\documentclass[reprint,showpacs,aps,flushbottom,nofootinbib,pra]{revtex4-1}

\usepackage{amsmath, amssymb, graphicx, psfrag, pstool, bm, color, hyperref}

\hypersetup{
    colorlinks=true,       % false: boxed links; true: colored links
    linkcolor=blue,          % color of internal links (change box color with linkbordercolor)
    citecolor=blue,        % color of links to bibliography
    filecolor=blue,      % color of file links
    urlcolor=blue           % color of external links
}

\usepackage{txfonts}

\definecolor{deepblue}{RGB}{0,0,128}
\definecolor{deepred}{RGB}{128,0,0}
\definecolor{deepcyan}{RGB}{0,128,128}

\newcommand{\bq}{\begin{equation}}
\newcommand{\eq}{\end{equation}}
\newcommand{\bn}{\begin{eqnarray}}
\newcommand{\en}{\end{eqnarray}}

\newcommand{\ket}[1]{\left|#1\right>}

\def\nm{\mbox{ nm}}

\def\G{\mbox{ G}}
\def\mG{\mbox{ mG}}

\def\uG{\mbox{ $\mu$G}}

\def\Gcm2{\mbox{ G$\times$cm$^2$}}

\def\us{\mbox{ $\mu$s}}
\def\ns{\mbox{ ns}}

\def\Hz{\mbox{ Hz}}
\def\kHz{\mbox{ kHz}}
\def\MHz{\mbox{ MHz}}
\def\GHz{\mbox{ GHz}}

\def\ea0{\mbox{ $ea_0$}}

\def\degC{\mbox{$^\circ$C}}

\def\V{\mbox{ V}}

\def\uA{\mbox{ $\mu$A}}

\def\Vcm{\mbox{ V/cm}}

\def\ecm{~\mbox{e$\cdot$cm}}

\begin{document}
\title{Search for the electron electric dipole moment using $\Omega$-doublet levels in PbO$^*$}

\author{S. Eckel}
\altaffiliation{Present address: National Institute of Standards and Technology, 100 Bureau Dr., Gaithersburg, MD 20899-8424}
\email{stephen.eckel@aya.yale.edu}
\author{P. Hamilton}
\altaffiliation{Present address: Department of Physics, University of California, Berkeley, California 94720, USA}
\author{E. Kirilov}
\altaffiliation{Present address: Universität Innsbruck, Institut für Experimentalphysik Technikerstrasse 25/4 A-6020 Innsbruck,Austria}
\author{H.W. Smith}
\author{D. DeMille}
\email{david.demille@yale.edu}

\affiliation{Yale University, Department of Physics, P.O. Box 208120, New Haven, CT 06520-8120}

\date{\today}

\pacs{14.60.Cd, 33.40.+f, 11.30.Er, 33.57+c}

\begin{abstract}
We present results of the first experiment to probe for the electric dipole moment (EDM) of the electron using an $\Omega$-doublet state in a polar molecule.  If the molecule is both massive and has a large molecular-fixed frame dipole moment, then the $\Omega$-doublet states have the potential to greatly increase the sensitivity of experiments searching for the EDM while also allowing for new methods of systematic error rejection.   Here, we use the metastable $a(1)^3\Sigma^+$ state of lead monoxide (PbO) to probe for the electron EDM.  Our best fit for the electron EDM of $d_e = (-4.4\pm9.5_\text{stat}\pm1.8_\text{syst})\times10^{-27}\ecm$ allows us to place an upper limit on the magnitude of the EDM of $|d_e|<1.7\times10^{-26}\ecm$ (90\% confidence).  While this is less stringent than limits from other, previous experiments, our work emphasizes the systematic error rejection properties associated with the $\Omega$-doublet level structure.  The results should inform the work of other, ongoing experiments that use molecules with analogous level structure.
\end{abstract}

\maketitle

\section{Introduction}
\label{sec:introduction}
The detection of a permanent electric dipole moment (EDM) of a fundamental particle would provide evidence of violation of charge-parity (CP) symmetry~\cite{Schiff1963}.  If an EDM exists, it violates both parity (P) and time (T) reversal symmetries.  Through the CPT theorem, which states that the combined operations of parity, time, and charge-conjugation (C) must be conserved in any Lorentz-invariant theory~\cite{Schwinger1951}, a violation of T symmetry, such as that by an EDM, is equivalent to violation of CP symmetry.  CP violation was first observed in the decay of the neutral kaon~\cite{Christenson1964}, and such violation can be explained through the standard model (SM).  However, the SM does not contain enough CP violation to explain the current matter-antimatter asymmetry in the universe~\cite{Balazs2005}.  Theories that go beyond the standard model generally provide for more CP violation, and therefore, larger EDMs~\cite{Pospelov2005,Booth1993,Khirplovich1991,Bernreuther1991,Hoogeveen1990}.

No EDM of a fundamental particle has yet to be detected; however, a recent experiment using YbF molecules has set the current experimental limit on the electron EDM ($e$EDM) of $|d_e|<1.05\times10^{-27}\ecm$ (90\% confidence)~\cite{Hudson2011}.  This experiment, like most of its predecessors~\cite{Regan2002,Griffith2009}, attempted to detect the $e$EDM by detecting a change in Larmor precession in the presence of an electric field.  To understand how these experiments work, consider a {\it free} electron in the presence of both electric and magnetic fields.  If the electron possesses an EDM, the magnitude of the resulting Larmor precession frequency vector $\bm{\omega}$ is given by
\begin{equation}
 \label{eq:basic_larmor_precession}
 \omega = \frac{1}{\hbar}|2 g_e \mu_B \mathbf{B} + 2 d_e \bm{\mathcal{E}}|\ ,
\end{equation}
where $\bm{\mathcal{E}}$ is the applied electric field, $d_e$ is the $e$EDM, $\mu_B$ is the Bohr magneton, $g_e$ is the g-factor for the electron, $\mathbf{B}$ is the magnetic field, and $\hbar$ is Planck's constant divided by 2$\pi$\footnote{We note that the precession frequency is a vector in the most general case.  While the experiment discussed in this work only measures the magnitude of the precession frequency vector, there are experiments that can measure not just the magnitude, but the direction of the precession.  The implications of such a measurement are discussed briefly in Appendix~\ref{app:phase_comb}.}.  If the two fields are nominally applied along the same axis and if the EDM is parallel to the magnetic dipole moment\footnote{I.e., both are anti-parallel to the spin.  Such a condition would imply $d_e<0$, just as the gyromagnetic ratio $\gamma_e = g_e\mu_B$ obeys $\gamma_e<0$.}, the magnitude of the precession frequency will be larger (smaller) if the two fields are parallel (anti-parallel).  To isolate the precession due to $d_e\mathcal{E}$, one typically measures the difference in the precession frequency when the relative direction of the $\mathcal{E}$ and $\mathbf{B}$ fields are reversed.

Because the electron is charged, it is not feasible to measure the precession frequency of free electrons, and  therefore atoms and molecules with unpaired electron spins are commonly used~\cite{KhriplovichLamoreaux,Commins2010}.  If an electric field is applied to such an atom or molecule, Eq.~\ref{eq:basic_larmor_precession} can still be applied, except $g_e$ is replaced by the $g$-factor of the bound state with which the measurement is performed, and $\mathcal{E}$ is replaced by an effective electric field $\mathcal{E}_{eff}$ with which the electron spin interacts.  In the non-relativistic limit, $\mathcal{E}_{eff}$ is equal to zero (a result known as the Schiff theorem), but due to relativistic effects can be non-zero or even larger than the applied electric field.  Atoms or molecules with a heavy nucleus can offer effective electric fields as high as 100~GV/cm.  Because $d_e<10^{-27}\ecm$, the shift in Larmor precession frequency will still be small, of order $2\pi\times0.1\Hz$.  The bias magnetic field $B$ must still be large enough to observe precession; in practice, $B$ is ordinarily of the order of $10$ to $10^3$~mG, corresponding to overall precession frequencies in the range of $\omega \approx 2\pi\times (10^4)$ to $2\pi\times (10^6)$~Hz.  Therefore, the $e$EDM may produce a tiny fractional shift of between $10^{-4}$ and $10^{-8}$ in the spin precession.

In any experiment where the expected signal represents such a small fractional shift, the experimentalist must take care to isolate systematic effects that mimic the sought-for shift.  For $e$EDM experiments, a magnetic field that is generated by the electric field, such as one created by leakage current that flows from one electrode to another, represents one of these possible systematic effects.

In experiments where the Larmor precession frequency is given by Eq.~\ref{eq:basic_larmor_precession}, reversal of the magnetic and electric fields is the primary method used to measure and reject systematic effects.  In this method, magnetic and electric fields are applied along a chosen laboratory axis and their directions relative  to that axis are reversed.  Four Larmor precession frequencies corresponding to the four possible applied field configurations are then measured.  Combinations of these measured frequencies then yield information regarding the $e$EDM and the applied fields.  For example, if $\omega_{i,j}$ represents the measured precession frequency when $B$ has sign $i$ and $\mathcal{E}$ has sign $j$, the sum of all four obeys $\omega_{+,+} + \omega_{+,-} + \omega_{-,+} + \omega_{-,-} = 8g\mu_B B_{av}$, where $B_{av}$ is the average magnetic field magnitude.  The $e$EDM can be found by forming the combination $\omega_{+,+} - \omega_{+,-} - \omega_{-,+} + \omega_{-,-}$, which assuming all the field magnitudes are identical, yields $8d_e\mathcal{E}_{eff}$.  Given that there are four different measured frequencies, there are two other {\it field parameters}, i.e., parameters that specify the applied fields.  These correspond to a non-reversing component of the magnetic field $B_{nr}$ and a change in the magnetic field magnitude $B_{corr}$ that is correlated with the {\it absolute} direction of the electric field.  However, a magnetic field component whose sign depends on both the sign of $\mathcal{E}$ and $B$, e.g., a magnetic field that is generated by leakage current that flows between the two electrodes, will be completely indistinguishable from  $e$EDM in this method.  In addition, no information regarding the electric field or the alignment of the two fields can be obtained solely from the measured precession frequencies.

\begin{figure}
 \center
 \includegraphics{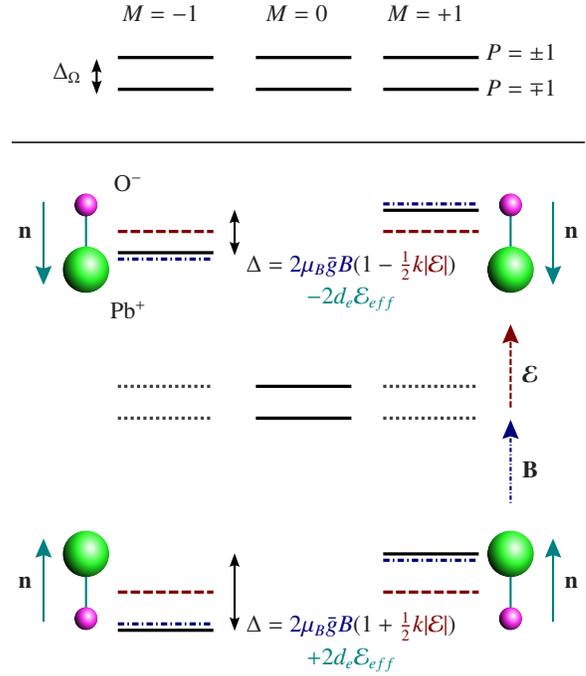}
 \caption{\label{fig:energy_structure} (Color online) Schematic of the energy level structure of a $J=1$ $\Omega$-doublet in a polar molecule.  (Top panel) With no applied electric or magnetic fields, there are two sets of three degenerate states, corresponding to the parity eigenstates $P$ and the three projections $M$ of the angular momentum $J$ along the quantization axis, respectively.  Which of the parity eigenstates has higher energy depends on the exact structure of the molecule in question; in the a(1)$^3\Sigma^+$ state of PbO used in this work, the $P=+1$ state lies lower in energy.  (Bottom panel) An applied electric field $\bm{\mathcal{E}}$ (red, long dash vector) mixes the opposite-parity states (shown by the dotted, gray lines) and induces a Stark shift/splitting between them.  The Stark-shifted levels (shown by the red, long dashed lines) have a net average molecular dipole moment $\mathbf{n} = \mu_a\hat{n}$, where $\hat{n}$ is the molecular axis, oriented either along or against the electric field.  Hence the field $\bm{\mathcal{E}}$ naturally defines the quantization axis.  An applied magnetic field $\mathbf{B}$ (blue, dash-dot vector) in the same direction as $\bm{\mathcal{E}}$ induces a Zeeman shift between the $M=\pm1$ states (shown by the blue, dash-dot lines).  The $e$EDM interacts with the effective electric field of the molecule, which has opposite sign for the two $N$ states, and modifies the Zeeman shift (black lines).  Last, the electric field causes mixing between the $J=1$ and $J=2$ states that effectively changes the $g$ factors of the upper and lower $N$-states by an amount $\frac{1}{2}k|\mathcal{E}|$.}
\end{figure}

Recently, considerable attention has turned to polar molecules with $\Omega$-doublet substructure~\cite{Meyer2006,Lee2009,Vutha2010,Loh2011,Grau2012}, because such systems have the possibility to reject systematics better than the simple atomic or molecular experiments described above~\cite{DeMille2001}.  Such $\Omega$-doublet substructure occurs generically in molecular states with internal electronic angular momentum $J_e \ge 1$.  In addition to having states with $\Omega$-doublet substructure, the molecular species must generally have one heavy nucleus,  two valence electrons in a triplet state, and at least one valence electron in a $\sigma$ or $\pi_{1/2}$ orbital in order that $\mathcal{E}_{eff}$ for the $\Omega$-doublet states is also large.  As shown in Fig.~\ref{fig:energy_structure}, it is possible to prepare such a molecule in quantum states that have opposite signs of the molecular polarization and hence opposite signs of the effective electric field.  This additional degree of freedom can possibly yield more information regarding experimental conditions such as field alignments, field magnitudes and leakage currents\footnote{Other potential systems are being studied for $e$EDM searches as well, including solid state systems~\cite{Eckel2012,Rushchanskii2010} and Cs and Rb atoms in an optical lattice~\cite{Weiss2003}.}.

A detailed description of the level structure of these states is given in Refs.~\cite{Bickman2009,Hamilton2010}, and a short review will be presented here.  With no applied magnetic and electric fields, the total angular momentum $J=1$ state has two degenerate manifolds of three states.  Each of the three states corresponds to a different projection $M$ of the total angular momentum along the laboratory $\hat{z}$ axis, as shown in the top panel of Fig.~\ref{fig:energy_structure}.  The two sets of degenerate states correspond to eigenstates\footnote{These parity states are best expressed in the basis $\ket{J,M,|\Omega| = 1,P = \pm1}$, where $\Omega$ is the projection of the electronic angular momentum $J_e$ along the molecular axis $\hat{n}$, i.e., $\Omega = \mathbf{J}_e\cdot\hat{n}$.  The parity states can be expressed in the signed $\Omega$ basis through $\ket{J,M,|\Omega| = 1,P = \pm1} = \frac{1}{\sqrt{2}}(\ket{J,M,\Omega = 1} - (-1)^J P\ket{J,M,\Omega = -1})$.} of the parity operator with quantum number $P=\pm1$.  Because of a Coriolis coupling to the total electronic angular momentum $\mathbf{J}_e$, these two parity states have a small difference of energy $\Delta_\Omega$.  Which of the two parity eigenstates has higher energy depends on the specific molecule and state.  For the purposes of this discussion, the state with higher energy will be denoted by $H$, and the state with lower energy will be denoted by $L$.  Note that while in the molecule-fixed frame there is an electric dipole moment $\mu_a$, the expectation value in the laboratory frame for either of these states is identically zero.  In addition, the $H$ and $L$ states have slightly different magnetic $g$-factors, which will be denoted as $g_H$ and $g_L$, respectively.

An applied electric field causes the molecules to polarize along or against the electric field.  The tensor Stark shift that accompanies this polarization is shown schematically in Fig.~\ref{fig:energy_structure}.  In the presence of $\bm{\mathcal{E}}$, the projection of the molecular-fixed dipole moment $\mu_a$ along the electric field, given by $N \equiv \text{sign}(\langle\bm{\mu_a}\rangle \cdot \bm{\mathcal{E}})$, becomes a good quantum number.  Thus, the direction of the electric field defines the quantization axis for the molecule\footnote{For this configuration, it is useful to define the basis $\ket{J,M,N} = \ket{J,M,\Omega = N\text{sign}(M)} = \frac{1}{\sqrt{2}}(\ket{J,M,|\Omega| = 1,P=+1} - (-1)^\Omega \ket{J,M,|\Omega| = 1,P=-1})$.}.  When a magnetic field is applied along the quantization axis, it interacts with the unpaired spins and causes a splitting between the $M=\pm1$ sublevels of $2g_{N_\pm}\mu_B B$, where $g_{N_+(N_-)}$ is the g-factor for the $N=1$ ($N=-1$) state.  As described in Sec.~VII of Ref.~\cite{Bickman2009}, the $g$ factors diverge when strongly polarized by an applied electric field, i.e.,
\bq
 \label{eq:dgogvsE}
 g_{N_+} - g_{N_-} \equiv \Delta g \approx \frac{3 \mu_a |\mathcal{E}|}{20B_r}(g_H+g_L)\ ,
\eq
where $B_r$ is the rotational constant.

The $e$EDM itself interacts with the effective electric field of the molecule, not the external field.   Because the states have different signs of $N$, we write the energy shift between the $M=\pm1$ states due to the $e$EDM as $E=-2\mathbf{d_e}\cdot\bm{\mathcal{E}}_{eff}$, where $\bm{\mathcal{E}}_{eff} = -\mathcal{E}_{eff}\hat{n}$ is the effective electric field with which the $e$EDM interacts; this quantity has opposite sign for the two $N$ states.  Once the applied electric field is sufficiently large to polarize the molecule, $\mathcal{E}_{eff}$ reaches its maximum value and becomes independent of the applied electric field\footnote{In the case of a paramagnetic atom, $\mathcal{E}_{eff}$ is linearly proportional to the applied electric field for all fields achievable in the laboratory.  This difference between atoms and molecules is notable and could possibly be used as an additional check for systematic effects.}.

Because a polarized molecule in the $J=1$ state has a large energy difference between the $M=0$ and $M=\pm1$ states, magnetic fields perpendicular to the quantization axis, defined by the electric field $\bm{\mathcal{E}}$, will have a minimal impact on the precession frequency.  Such a transverse magnetic field $B_T$ will couple states with $M$ and $M'=M\pm1$, and therefore there is no shift to first order in $B_T$.  To second order, assuming a fully polarized molecule and neglecting the effect of the $J=2$ states, the shift in energy $\delta E$ due to $B_T$ for the four $N=\pm1$, $M=\pm1$ states is given by
\begin{eqnarray}
 \label{eq:transverseB_energy}
 \delta E_{N,M} & = & -\frac{\mu_B^2B_T^2}{2\mu_a\mathcal{E}}\left[(g_H^2+g_L^2)\left(N+M\frac{2\bar{g}\mu_B B_z}{\mu_a\mathcal{E}}\right) \right. \nonumber \\
 & & \hspace{0.25in} \left. - (g_H^2-g_L^2)\frac{\Delta_\Omega}{2\mu_a\mathcal{E}}\right]\ ,
\end{eqnarray}
where $B_z$ is the magnetic field along the electric field, $B_T$ is the transverse magnetic field perpendicular to it and $\bar{g} = \frac{1}{2}(g_H+g_L)$.  Therefore, while a transverse field may lead to a shift in the overall energy of the $M=\pm1$ states, such a shift is mostly common mode.  The actual change in the precession frequency for the state with quantum number $N$ due to the transverse magnetic field $\delta\omega_N^T$ is given by
\begin{equation}
 \label{eq:transverseB_larmor}
 \delta \omega_N^T = \frac{\delta E_{N,M=+1} - \delta E_{N,M=-1}}{\hbar} = -2(g_H^2+g_L^2)\frac{\bar{g}\mu_B^3}{\hbar}\frac{B_z B_T^2}{(\mu_a\mathcal{E})^2}\ ,
\end{equation}
which is the same for both $N=\pm1$.

Collecting all the terms above, we can write the magnitude of the precession frequency of the $N=\pm1$ states to lowest order in the various fields as
\begin{eqnarray}
 \omega_N & = & \frac{|E_{N,M=+1}-E_{N,M=-1}|}{\hbar} \nonumber \\
 & = & \frac{1}{\hbar}\left|2\bar{g}\mu_B (\mathbf{B}\cdot \hat{\mathcal{E}})\left(1 \vphantom{\frac{(\hat{E})^2}{\mathcal{E}^2}} + \frac{N k|\bm{\mathcal{E}}|}{2} \right.\right.  \nonumber \\
 & & \hspace{0.25in} \left.\left. - (g_H^2+g_L^2)\frac{\mu_B^2(\mathbf{B}\times\hat{\mathcal{E}})^2}{(\mu_a\mathcal{E})^2}\right) + 2 N d_e\mathcal{E}_{eff}\right| \label{eq:larmorprec} \ ,
\end{eqnarray}
where $k= 3 \mu_a/10B_r$.  Unlike experiments based upon Eq.~\ref{eq:basic_larmor_precession}, it is now possible to measure eight different Larmor precession frequencies corresponding to the four different applied field configurations and two different N states.  These eight measured frequencies are denoted by $\omega_{i,j,k}$, where $i=\mbox{sgn}(N)$, $j=\mbox{sgn}(\mathcal{E})$ and $k = \mbox{sgn}(B)$, and the $e$EDM can be found by taking the combination of the measured frequencies that is odd in $N$, $\mathcal{E}$, and $B$, i.e., $\sum_{ijk} ijk \omega_{i,j,k} = 16d_e\mathcal{E}_{eff}$. 

Moreover, an experiment based on measuring the eight different Larmor frequencies given by Eq.~\ref{eq:larmorprec} can yield more information regarding the applied fields.  First, with applied electric fields $\mathcal{E}\lesssim 100\Vcm$, $k\mathcal{E}$ is typically of the order of $10^{-2}$.  Therefore, the two $N$ states respond almost equally to the applied magnetic field and act as an internal comagnetometer, allowing cancellation of the effects of a fluctuating or systematically changing magnetic field.  For example, this can be used to distinguish frequency shifts due to a field generated by current flowing between the electrodes from shifts due to the $e$EDM.  Second, while the electric field dependence of the $g$ factors prevents perfect cancellation of magnetic field effects, it does allow extraction of information regarding the electric field.  This information can be used to determine whether the electric field magnitude is changing with time or upon reversal of its direction.

An experiment based upon Eq.~\ref{eq:larmorprec} also can reject systematic errors due to misalignment of the fields.  With typical alignment errors between the electric and magnetic fields, the ratio $\mu_B \mathbf{B}\times\hat{\mathcal{E}}/(\mu_a\mathcal{E})$ can be of the order of $10^{-5}$, leading to rejection of transverse magnetic fields of the order of $10^{-10}$.  Moreover, motional magnetic fields generated by the relativistic $\mathbf{v}\times\bm{\mathcal{E}}$ effect will have a negligibly small impact on the $e$EDM measurement.

In this work, we examine the systematic rejection properties of an eEDM search based on this level structure, using an experiment based on the a$^3\Sigma^+$ state of lead monoxide (PbO).  PbO has proven interesting for an $e$EDM search for a multitude of reasons.  First, it can be produced with high-densities in a vapor cell~\cite{DeMille2000,Hunter2002,Kozlov2002,Kawall2004,Bickman2009}.  Second, it has both large electric dipole moment ($\mu_a/\hbar\approx2\pi\times1.64$~MHz V$^{-1}$cm$^{-1}$~\cite{Hunter2002}) and a large effective electric field ($\mathcal{E}_{eff}\sim25$~GV/cm~\cite{Kozlov2002,Petrov2005}).  While the state is sensitive to magnetic fields, with $\bar{g}\approx0.86$~\cite{Kawall2004}, the difference in the $g$ factors with magnetic fields obeys $k\approx7\times10^{-5}$~V$^{-1}$cm$^{-1}$, where we have used $B_r/\hbar = 2\pi\times(7.054\GHz)$~\cite{Martin1988}.  This value of $k$ ensures that the molecule will behave as a good comagnetometer for our values of applied electric fields.

Although the present work was not able to establish a new limit on the $e$EDM, many future experiments depend on the capabilities for the systematic rejection of the level structure shown in Fig.~\ref{fig:energy_structure}, and we believe the findings of this work will prove useful for those experiments.  We first discuss the apparatus\footnote{For a comprehensive description of the appartus and basic experimental procedure used, see also Ref.~\cite{Hamilton2010}.} used in this study in Sec.~\ref{sec:apparatus}.  The suppression of effects due to transverse fields is shown experimentally in Sec.~\ref{sec:transverse_fields}.  A comprehensive list of frequencies or phases measured in such an experiment, along with a formal definition of the measurable systematics and physical quantities, will be developed in Sec.~\ref{sec:systematic_rejection}.  In addition to the effects detailed in Sec.~\ref{sec:systematic_rejection}, we consider the effect of field gradients on the experiment in Sec.~\ref{sec:gradients}; these prove to be the dominant source of systematic errors in this work.  The final result of the experiment is detailed in Sec.~\ref{sec:edm_limit}.  Note that for the remainder of this article, we shall take $\hbar=1$ and treat energy and angular frequency as identical quantities.

\section{Apparatus}
\label{sec:apparatus}

\begin{figure}
 \center
 \includegraphics{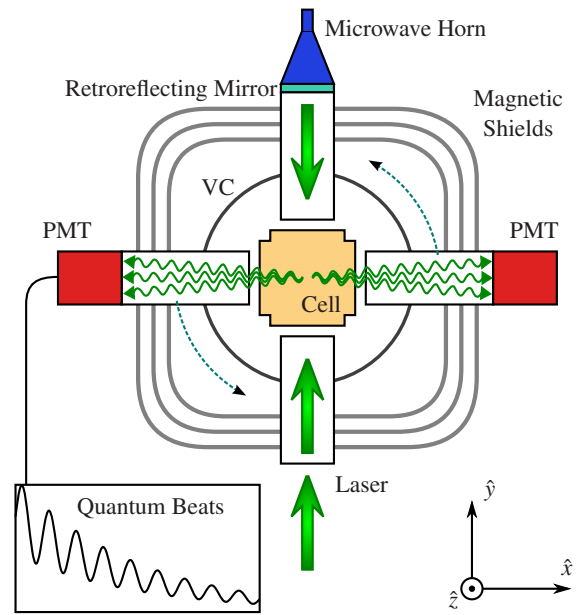}
 \caption{\label{fig:schematic} (Color online) A schematic of the experiment.  The cell is placed at the center of vacuum chamber (VC).  Light from a laser enters through a quartz light-pipe to illuminate the molecules, and the resulting quantum beat fluorescence is observed using photomultiplier tubes (PMTs).  Microwaves can also be applied to the molecules through a microwave horn.   The vacuum chamber and cell are contained within three layers of magnetic shielding.  The magnetic field coils and the oven that heats the vapor cell are not shown.  The green, dotted arrows indicate the direction of rotation of the resulting quantum beat fluorescence.}
\end{figure}

For our experiments, a gas of PbO molecules is created in a high-temperature vapor cell, as shown in Fig.~\ref{fig:schematic}.  The cell is shaped roughly like a cube with approximately 3~in. sides and is made from fused alumina, which forms the walls and structural supports.  Four,  2~in.-diameter, yttrium-aluminum-garnet (YAG) windows provide optical access from four sides of the cell.  Re-entrant electrodes, with an approximate diameter of 2.25~in., protrude into the top and bottom with a spacing of 1.5~in.  These electrodes are made from gold foil which has been adhered onto a flat, sapphire substrate using oxide bonding~\cite{Bailey1978}.  Surrounding the main electrodes are guard ring electrodes of inner diameter 2.5~in. and outer diameter 2.75~in.  When a larger voltage is applied to these guard rings than the main electrodes, the electric field becomes more uniform in the main volume of the cell.

An oven heats the vapor cell to $\sim\!700\degC$, where the partial pressure of PbO is $\sim\!10^{-4}$~Torr and the total pressure of PbO vapor (dominated either by Pb$_4$O$_4$ or PbO) is $\sim\!10^{-3}$~Torr~\cite{Lopatina2006,Popovic1997}.  The heating elements are formed from laser-cut tantalum foil, designed to minimize the overall self-inductance, and are held in place by a quartz structure.  Because tantalum will oxidize when heated in atmosphere, the oven and vapor cell are placed within an aluminum vacuum chamber with 18~in. height and 14~in. diameter, where pressures of approximately $10^{-5}$~Torr are achieved.  Two-inch diameter quartz light pipes protrude through the walls of the vacuum chamber and provide the necessary optical access for excitation and detection of the molecules.

The vacuum chamber is surrounded by multiple magnetic field coils.  A 300 turn, 10.5~in. radius Helmholtz coil generates a magnetic field in the $\hat{z}$ direction.  Cosine-type coils apply uniform fields in the $\hat{x}$ direction and $\hat{y}$ direction.  A set of gradient coils can generate all possible linear magnetic field gradients.  The vacuum chamber and magnetic field coils are placed within three layers of $\mu$-metal shielding, which provide a shielding factor of the order of $10^3$.

A short laser pulse of 548~nm light prepares the molecules into the a$^3\Sigma^+$ state of PbO.  The pulse is generated by a narrowband, continuous-wave-seeded, pulse-pumped dye amplifier~\cite{Schwettmann2007,Drell1979}.  A diode laser, which is amplified using a semiconductor tapered amplifier and frequency-doubled using a periodically-poled lithium niobate (PPLN) waveguide, generates the seed laser light.  An injection-seeded, frequency-doubled Nd:YAG laser with an output pulse of approximately 3~ns in duration and a repetition rate of 100~Hz acts as the pump for Fluoresin 548 dye.  The output energy of the full dye amplifier system has an energy of approximately 7--10~mJ per pulse.  Because the dye amplifier's linewidth ($\lesssim200$~MHz) is smaller than the Doppler width of the molecular transition at $T \approx 1000$~K ($\Gamma_D \approx 800\MHz$), a mirror retroreflects the beam to excite molecules in a broader range of velocity classes. 

With $\hat{x}$-polarized light, the laser drives the transition from the absolute ground state X$^1\Sigma(v=0,J=0)$ into the manifold of $^3\Sigma^+(v=5,J=1)$ sublevels and coherently populates the $N=\pm1$, $M=\pm1$ levels shown in Fig.~\ref{fig:energy_structure}.  Neither the $M= \pm 1$ Zeeman splitting nor the $N=\pm 1$ Stark splitting is resolved within the Doppler width of the transition.  Because of the non-negligible electric field inhomogeneity ($\sim\!1\%$), any coherence between the $N=\pm1$ states is quickly lost\footnote{We estimate that the $N=\pm 1$ coherence is lost in less than $250\ns$.}.  The resulting fluorescence signal indicates that the $a$ state has an effective lifetime of approximately $\tau_a \approx 50\us$ (see below for details), with decay presumed to be due in nearly equal parts to spontaneous emission and quenching on cell walls.  As the molecules decay to the ground state, quantum beats are created by interference of the decay paths to $M=0$ ground state sublevels~\cite{Bickman2009}.  The modulated fluorescence signal is detected using two photomultiplier tubes (PMTs) that are oriented along the $\hat{x}$-axis, which is perpendicular to the laser's direction of propagation.   Two filters mounted in front of each PMT, a KG-4 infrared-blocking colored glass and a custom $554\pm104\nm$ optical interference filter, serve to block the primary spectrum of the substantial blackbody radiation present at $T\approx 1000$~K.  The quantum beat fluorescence along the $x$-direction is polarized along the $\hat{y}$ axis; therefore, wire grid polarizers are also used to filter out other background light.
 
In order to resolve the signal from the two different $N$ states, one of two techniques is used.  As described in Sec. X of Ref.~\cite{Bickman2009}, microwaves that are resonant with either the $\ket{J=1,M=\pm1,N=-1} \rightarrow \ket{J=2,M=\pm2,N=-1}$ or $\ket{J=1,M=\pm1,N=+1} \rightarrow \ket{J=2,M=\pm2,N=+1}$ are applied to the molecules for a duration $\tau_M\gg1/\Omega_R$, where $\Omega_R$ is the Rabi frequency of the microwave drive.  This causes decoherence of the beat signal from the particular $N=\pm1$ state with which the microwaves are resonant.  The frequencies of these transitions are approximately 28~GHz.  The microwaves are generated using a custom-built microwave source, and are applied to the molecules via a microwave horn.

\begin{figure}
 % Source of this figure is 2011-07-13/2054/step_0314.bin (chosen at random from the run).
 \center
 \includegraphics{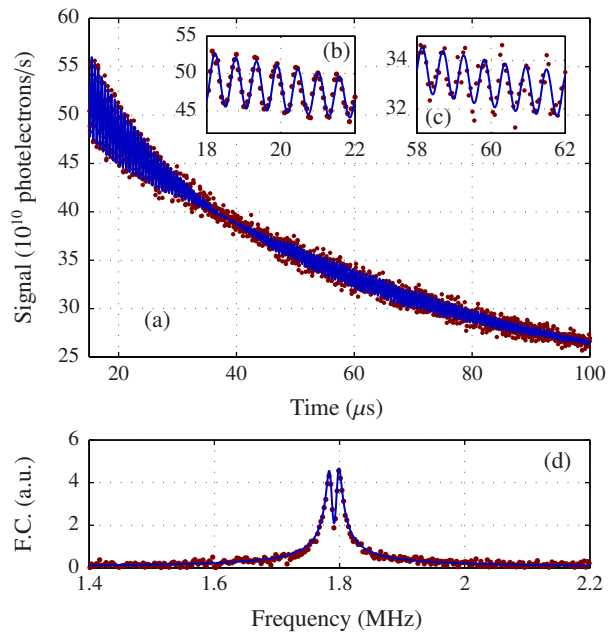}
 \caption{\label{fig:two_beat_example} (Color online) An example of the resulting two beat signal from a single laser shot.  At sufficiently large applied magnetic and electric fields, the difference in the $g$ factors coupled with the size of the average beat frequency allows for resolution of both $\Omega$-doublet states at the same time.  (a) The two precession frequencies generate a beat note in the raw signal (shown as red points), as evidenced by the node near $t=38\us$.  This data fits well to Eq.~\ref{eq:two_beat_fit} (shown as a blue curve).  (b) At the beginning of the decay, oscillations at the average of the two frequencies are seen.  (c) The beat signal still fits to Eq.~\ref{eq:two_beat_fit} after the node, where the phase of the signal has shifted by $\pi$. (d) A discrete Fourier transform with a rectangular window function of both the data (shown as red points) and the fit (shown as a blue curve) shows two resolved frequencies in the spectrum of the quantum beat signal.}
\end{figure}

Alternatively, one can also resolve the signals of the two $N$ states by applying large magnetic and electric fields.  If both fields are sufficiently large, the difference in the precession frequency between the two $N$ states can become larger than $1/T_2^*$, where $T_2^*$ is the lifetime of the quantum beats.  In this case, both $N$ states contribute to the quantum beat signal, and the two frequency components can be resolved, as shown in Fig.~\ref{fig:two_beat_example}.  In this case, there is no loss of signal due to microwave preparation, and therefore the sensitivity of the experiment to the $e$EDM is generally better in this two-beat case than in the microwave-erasure technique described above.

The resulting quantum beats are fit to the function
\begin{equation}
 \label{eq:single_beat_fit}
 S(t) = b(t) + e^{-\Gamma t}a_1\cos(\omega_1 t + \phi_1)
\end{equation}
or
\begin{equation}
 \label{eq:two_beat_fit}
 S(t) = b(t) + e^{-\Gamma t}[a_1\cos(\omega_1 t + \phi_1) + a_2\cos(\omega_2 t + \phi_2)]\ ,
\end{equation}
for microwave-erasure and two-beat techniques, respectively.  In these fits, $a_i$, $\phi_i$, $\omega_i$ and $\Gamma$ are free parameters, $t$ is the time since the laser pulse, and $b(t)$ is the background signal.  This background signal is determined directly from the data by using a low-pass, zero-phase-shift filter (a digital filter that moves both forward and backward through time~\cite{Oppenheim2010}).  In general, $b(t)$ can be fit to
\begin{equation}
 \label{eq:background_fit}
 b(t) = \mathcal{A}_A e^{-t/\tau_A} +  \mathcal{A}_a e^{-t/\tau_a} + b_{bb}\ ,
\end{equation}
where $\mathcal{A}_i$, $\tau_i$ and $b_{bb}$ are tunable constants.  The constant term $b_{bb}$ allows for fitting the blackbody background.  The first, fast exponential term generally obeys $\tau_A\approx (1/\Gamma)\times10^{-1}$ and is believed to be fluorescence from molecules that were excited to the nearby A$^3\Pi$ electronic state.  The second, slower exponential decay obeys $\tau_a\lesssim1/\Gamma$ and is believed to be the fluorescence lifetime of molecules excited to the $a$ state that are not spin-polarized in the J=1 state and hence do not contribute to the quantum beat signal.  This fluorescence lifetime is the aforementioned $\tau_a\approx50\us$.

Using a vapor cell at $700\degC$ poses some unique challenges.  First, in order to maintain the temperature, current must flow through the heating elements.  Such current will generate magnetic fields that could interfere with the $e$EDM measurement.  For this reason, the current through the heaters is brought to zero before the laser pulse is fired and subsequently restored approximately $400\us$ after the laser pulse.  In order to avoid inducing eddy currents in the aluminum vacuum chamber, the heaters are supplied with a 10~kHz AC signal, which during 1~ms-long turn-on and turn-off periods is modulated by an envelope of the form $1-\sin^2(\xi t)$, where $\xi = 2\pi\times( 2\kHz)$~\cite{Jiang2008}.  Stereo audio amplifiers are driven with an arbitrary function generator to provide this AC signal with the $\sim\!1.1$~kW of power necessary to heat the vapor cell.

\begin{figure}
 % Source of this figure is 2011-07-19, leakage current measurements 1612 and 1659.
 \center
 \includegraphics{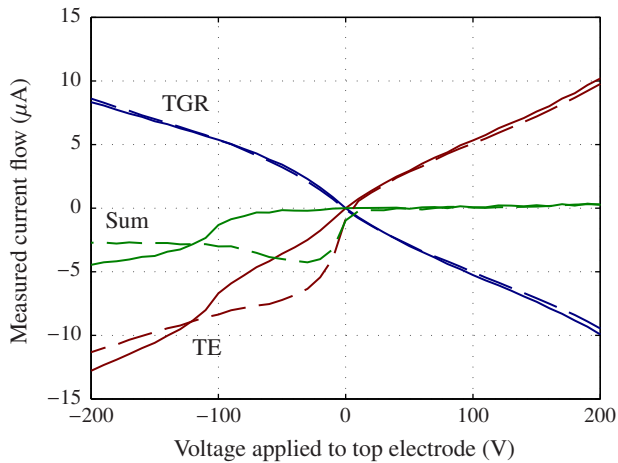}
 \caption{\label{fig:electron_emission} (Color online) Evidence of electron emission.   The current flowing into/out of the top electrode (TE) and its guard ring electrode (TGR) is shown as a function of the voltage applied to the TE, while the TGR is held at ground.  When the applied voltage is greater than zero, the current flow shows an approximately Ohmic-behavior, with current from the TE being detected flowing into the TGR such that the sum is zero.  When the voltage is negative, however, the sum is not zero and an excess of negative current is observed flowing from the top electrode.  Directly above the cell is a heating element that is not in physical contact with the vapor cell.  By biasing this heating element to $-100$~V  (solid lines) or $0\V$ (dashed lines), the threshold voltage at which excess current is observed is changed.  This process is analogous to the change in current flow provided by an electron-emitting filament and a biased grid electrode in a vacuum-tube triode.}
\end{figure}

Secondly, operating at $700\degC$ greatly decreases the electrical resistivity of most insulators, including the fused alumina and YAG used in our vapor cell.  While this lower resistivity generally creates large leakage currents, an even more insidious problem plagues the vapor cell.  At such large temperatures, electron emission is observed, as shown in Fig.~\ref{fig:electron_emission}.  This electron emission and associated voltage drops across the path between the electrode leads and the emission surface have the possibility to seriously distort the electric field, e.g., changing its magnitude and direction.  Making $e$EDM measurements in the presence of this problem provides one of the most rigorous tests of systematic quantification and rejection, and thus is integral to the primary goal of this work.

\section{Rejection of Transverse Magnetic Fields}
\label{sec:transverse_fields}

\begin{figure}
 % Source of this figure is 2012-05-09/positive_b_field_fits.sfit, Detector \#1, E<0, B>0.  The original data is from, 5/9/12, run 2047 (most likely, original source is confusing).
 \center
 \includegraphics{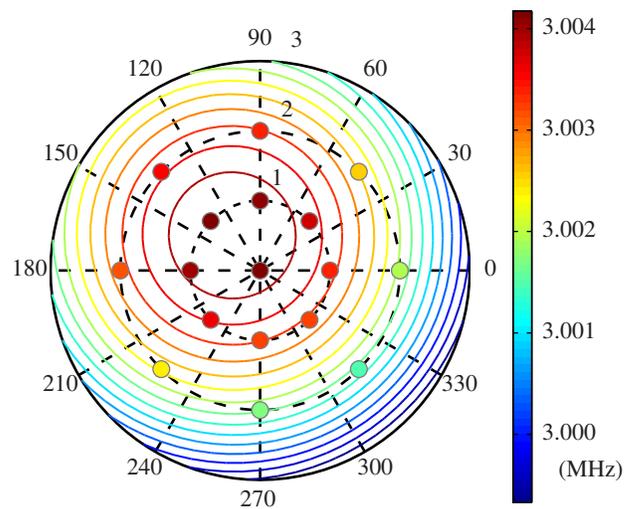}
 \caption{\label{fig:electric_field_angle} (Color online) Measurement of the alignment of the electric and magnetic fields.  In this polar projection plot, the radial and azimuthal components correspond to the polar angle and azimuthal angle, respectively, in a spherical coordinate system where $+\hat{z}$ is defined as the nominal direction of the magnetic field.  With a fixed electric field applied to the molecules, the applied magnetic field was tilted from the $\hat{z}$ axis without changing its magnitude, and the resulting average beat frequency $\omega_{av}$ was measured at the colored points shown.  The circular contours represent the best fit 2-D polynomial to the data.  The place where the beat frequency maximizes, at $\theta=(0.64\pm0.02)^\circ$ and $\phi = (128.2\pm0.1)^\circ$, represents the place of maximal alignment between the magnetic and electric fields.  The fractional uncertainty in the beat frequency measurement is approximately $10^{-5}$, using 512 laser shots per point.}
 % The radial and azimuthal components correspond to the polar angle and azimuthal angle, respectively, in a spherical coordinate system where $+\hat{z}$ is defined as the direction of the magnetic field when the field due only to the Helmholtz coil is applied and $\hat{x}$ is the direction defined by the magnetic field generated by the cosine coil most aligned with the axis of the detectors. 
\end{figure}

According to Eq.~\ref{eq:larmorprec}, the precession frequency is determined primarily by the projection of $\mathbf{B}$ onto the electric field $\bm{\mathcal{E}}$, which determines the quantization axis of the molecule.  Any transverse field only affects the precession frequency at higher order.  For a typical experiment, the condition $\mu_B(\mathbf{B}\times\hat{\mathcal{E}})\ll\mu_a\mathcal{E}$ is satisfied, and the relevant transverse-field term in Eq.~\ref{eq:larmorprec} can be neglected.  The precession frequency, averaged between both $N=\pm1$ states, can then be written as 
\bq
 \label{eq:avg_precfreq}
 \omega_{av} \equiv \frac{1}{2}\left(\omega_+ + \omega_-\right) = 2\bar{g}\mu_B(\mathbf{B}\cdot\hat{\mathcal{E}})\ ,
\eq
where $\omega_\pm$ is the precession frequency for the $N=\pm1$ state\footnote{For the present discussion, we do not reverse the magnetic or electric fields, so we omit the $j$ and $k$ indices present in Section~\ref{sec:introduction}.}.

If the alignment between the magnetic and electric fields is changed, Eq.~\ref{eq:avg_precfreq} implies that the average precession will change as well, even if the magnitude of $\mathbf{B}$ remains constant.  Such an effect is shown in Fig.~\ref{fig:electric_field_angle}.  Here, an electric field is applied to polarize the molecules.  The magnetic field is then tilted from its nominal direction without changing its magnitude.  (In this coordinate system, the vertical or $\hat{z}$ axis is defined by the magnetic field applied by the Helmholtz coil, and $\hat{x}$ is the direction defined by the magnetic field generated by the cosine coil most aligned with the axis of the detectors.)  If the angle between the two fields is denoted by $\gamma$, then Eq.~\ref{eq:avg_precfreq} becomes
\begin{eqnarray}
 \omega_{av} & = & 2\bar{g}\mu_B B \cos\gamma = 2 \bar{g}\mu_B B (\cos\theta\cos\theta_E \nonumber \\
 & & \hspace{0.7in} + \sin\theta\sin\theta_E\cos(\phi-\phi_E)\ ,
\end{eqnarray}
where $\theta_E$ and $\phi_E$ are the polar and azimuthal angles of the electric field, respectively, and $\theta$ and $\phi$ are the polar and azimuthal angles of the magnetic field, respectively.   If both polar angles are small, this can be expanded to second order and compared to a second-order polynomial fit to the data to extract the relative angles.  The best fit polynomial is shown as the contours in Fig.~\ref{fig:electric_field_angle}, and is accurate to within the errors of our calibration of the magnetic field.

The fit in Fig.~\ref{fig:electric_field_angle} not only justifies the approximation that misaligned components of the magnetic field are small compared to the primary component, but also demonstrates that the relative angles of the electric and magnetic fields in the experiment can be measured with much better than $1^\circ$ precision.

\section{Field Reversals and Frequency Combinations}
\label{sec:systematic_rejection}

As discussed in Sec.~\ref{sec:introduction}, there are eight different precession frequencies $\omega_{i,j,k}$ that can be measured in a molecule with $\Omega$-doublet substructure.   The indices $i=\mbox{sgn}(N)$, $j=\mbox{sgn}(\mathcal{E})$ and $k = \mbox{sgn}(B)$ specify the $N$ state measured and the field configuration relative to a chosen laboratory axis.  As in the case of an experiment based on Eq.~\ref{eq:basic_larmor_precession}, these eight different frequencies can be summed together to form a frequency {\it combination}.  Let us define a combination of measured frequencies $\omega_{i,j,k}$ as
\bn
 \Delta_{i,j,k}\omega & = & \omega_{+,+,+} + i\omega_{-,+,+} + j\omega_{+,-,+} + k\omega_{+,+,-} \\
     & & + ij\omega_{-,-,+} + ik\omega_{-,+,-} + jk\omega_{+,-,-} + ijk\omega_{-,-,-} \nonumber \ .
\en
Note that because the frequency shift due to the $e$EDM is odd in all three reversals [see Eq.~\ref{eq:larmorprec}], $\Delta_{-,-,-}\omega$ will correspond to the $e$EDM signal channel.  Also, while this discussion focuses on frequency measurements, the following discussion applies equally well to experiments that measure the phase precession of molecules in a beam.  The differences between the two types of experiments are detailed in Appendix~\ref{app:phase_comb}.

\subsection{N-even combinations}
To determine the physical meaning of each combination, let us first consider the combinations that are even under reversal of $N$: $\Delta_{+,+,+}\omega$, $\Delta_{+,-,+}\omega$, $\Delta_{+,+,-}\omega$, $\Delta_{+,-,-}\omega$.  Because the $g$-factors of the $N=\pm1$ states are affected by the electric field in the equal but opposite way, these combinations are nominally independent of the electric field applied.  For this reason, they give information only regarding the magnetic field.

There are four total electric and magnetic field configurations and four combinations that are even in $N$.  Let us examine each combination in turn.  First, the term $\Delta_{+,+,+}\omega$ is proportional to the precession frequency averaged over both $\Omega$-doublets and the four states of $\bm{\mathcal{E}}$ and $\mathbf{B}$.  Therefore, we define $\Delta_{+,+,+}\omega = 16\bar{g}\mu B_{av}$, where $B_{av}>0$ is the average magnitude of the magnetic field $\mathbf{B}$ projected onto $\bm{\mathcal{E}}$.  A non-zero $\Delta_{+,+,-}\omega$, corresponding to a shift of the precession frequency upon reversal of the magnetic field, can be due to a non-reversing magnetic field $B_{nr}$, i.e., $\Delta_{+,+,-}\omega = 16\bar{g}\mu_B B_{nr}$.  It is also possible that the electric field can produce a magnetic field whose direction is correlated with the direction of $\bm{\mathcal{E}}$ that will add vectorially to the applied magnetic field.  Depending on the direction of the applied magnetic field, this additional field may lead to an increase or a decrease in the overall magnitude of the magnetic field.  The average precession frequency will then change when both the electric and magnetic fields are reversed.  The most obvious source of this additional $\bm{\mathcal{E}}$-induced magnetic field is due to leakage currents; therefore, the associated frequency combination shall be denoted as $\Delta_{+,-,-}\omega = 16\bar{g} \mu_B B_{leak}$.  There is one additional combination, namely $\Delta_{+,-,+}\omega$.  A non-zero $\Delta_{+,-,+}\omega$ could arise if there is a change in the magnetic field magnitude that is directly correlated with the sign of the electric field, but not with the sign of magnetic field\footnote{This could arise, for example, due to unwanted electronic coupling between the magnetic and electric field power supplies, e.g., coupling due to a ground loop.}.  The associated frequency combination shall be denoted by $\Delta_{+,-,+} = 16\bar{g} \mu_B B_{corr}$.

Given the above field parameterization, we can now ask what is the measured magnetic field magnitude for a given field configuration?   For a particular field configuration, the measured, average frequencies between the $N=\pm1$ states ($\omega_{av,j,k} = \frac{1}{2}\sum_i\omega_{i,j,k}$) can be used to define the measured magnetic field magnitude $B_{meas,j,k}$ through the relationship $\omega_{av,j,k} = 2\bar{g}\mu_B B_{meas,j,k}$.  The relationships between the four magnetic field parameters ($B_{av}$, $B_{corr}$, $B_{nr}$, and $B_{leak}$) and $B_{meas,j,k}$ are shown in Table~\ref{tab:meas_Efield}.

\begin{figure}
 \center
 \includegraphics{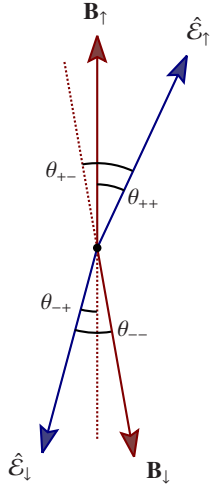}
 \caption{\label{fig:angles} (Color online) The angle interpretation for $B_{nr}$, $B_{corr}$, and $B_{leak}$.  The electric field direction is defined by two vectors $\hat{\mathcal{E}}_\uparrow$ and $\hat{\mathcal{E}}_\downarrow$ that correspond to it being in the positive and negative directions, respectively.  The magnetic field is defined by two vectors $\mathbf{B}_\uparrow$ (arbitrarily defined to be the $\hat{z}$ direction in this plot) and $\mathbf{B}_\downarrow$ that correspond to it being in the positive and negative directions, respectively.  If no direction is equal, there are four angles formed between the electric field axes and magnetic field axes.  These angles are specified by $\theta_{jk}$, where $j$ and $k$ correspond to the nominal sign of $\hat{\mathcal{E}}$ and $\mathbf{B}$, respectively, in the laboratory frame.}
\end{figure}

\begin{table}
 \center
 \begin{tabular}{ll|r|r|}
  $j$ & $k$ &  $B_{meas,j,k}$ & $\mathcal{E}_{meas,j,k}$ \\
  \hline
   $+$ & $+$ & $B_{av} + B_{corr} + B_{leak} + B_{nr}$ & $\mathcal{E}_{av} + \mathcal{E}_{corr} + \mathcal{E}_{nr}$ \\
   $-$ & $+$ & $B_{av} - B_{corr} - B_{leak} + B_{nr}$ & $\mathcal{E}_{av} + \mathcal{E}_{corr} - \mathcal{E}_{nr}$ \\
   $+$ & $-$ & $B_{av} + B_{corr} - B_{leak} - B_{nr}$ & $\mathcal{E}_{av} - \mathcal{E}_{corr} + \mathcal{E}_{nr}$ \\
   $-$ & $-$ & $B_{av} - B_{corr} + B_{leak} - B_{nr}$ & $\mathcal{E}_{av} - \mathcal{E}_{corr} - \mathcal{E}_{nr}$ \\
   \hline\hline
  \end{tabular}
 \caption{\label{tab:meas_Efield}  The parametrization of the measured magnetic field magnitude $B_{meas}$ and measured electric field magnitude $\mathcal{E}_{meas}$, for the four combinations of the applied fields.  The fields are nominally applied along a chosen laboratory axis, and $j = \mbox{sgn}(\mathcal{E})$ and $k=\mbox{sgn}(B)$ specify the direction relative to that axis.  For both the magnetic and electric fields, the average fields ($B_{av}$ and $\mathcal{E}_{av}$) are assumed to be both larger than zero and much larger than the magnitude of any other component.  Note that for the case of the electric field, there are three parameters and four field configurations.  The missing parameter has the same symmetry properties as the $e$EDM, and is therefore indistinguishable from it.}
\end{table}

Here we have parameterized the frequency combinations in terms of magnetic field components often discussed in the context of additional fields in a traditional $e$EDM experiment.  However, there is another way to understand what might cause non-zero values of $\Delta_{+,-,+}\omega$, $\Delta_{+,+,-}\omega$ or $\Delta_{+,-,-}\omega$.  As discussed in Sec.~\ref{sec:transverse_fields}, a change in precession frequency can occur if the axis of the magnetic field forms an angle with the electric field and this angle shifts upon reversal of either field.  To understand how the relative angle might generate a non-zero $\Delta_{+,-,+}\omega$, $\Delta_{+,+,-}\omega$ or $\Delta_{+,-,-}\omega$, consider the specific case shown in Fig.~\ref{fig:angles}.  Let the magnetic field in the positive (negative) state be defined as the vector $\mathbf{B}_{\uparrow(\downarrow)}$ and the direction of the electric field in its positive (negative) state be defined by the vector $\hat{\mathcal{E}}_{\uparrow(\downarrow)}$.  When both the magnetic and electric fields are in the positive state, the measured precession frequency averaged over the $N=\pm1$ states will be given by
\bq
 \omega_{av,+,+} = 2\bar{g}\mu_B|\mathbf{B}_\uparrow\cdot\hat{\mathcal{E}}_\uparrow| = 2\bar{g}\mu_B B_\uparrow\cos\theta_{++} \ ,
\eq
Likewise,
\bq
 \omega_{av,-,+} = 2\bar{g}\mu_B|\mathbf{B}_\uparrow\cdot\hat{\mathcal{E}}_\downarrow| = 2\bar{g}\mu_B B_\uparrow\cos\theta_{-+} \ .
\eq
Thus, there can be a change in the precession frequency if $\theta_{-+}\neq\theta_{++}$. If we assume the magnitude of the magnetic field remains unchanged when it is reversed, i.e., $B_\uparrow = B_\downarrow = B_{0}$, all of the magnetic field components defined above may be rewritten in terms of these angles:
\begin{subequations}
 \begin{eqnarray}
 B_{av} & = & \frac{1}{4} B_{0} (\cos\theta_{++} + \cos\theta_{-+} + \cos\theta_{+-} + \cos\theta_{--}) \\
 B_{corr} & = & \frac{1}{4} B_{0}(\cos\theta_{++} - \cos\theta_{-+} + \cos\theta_{+-} - \cos\theta_{--}) \\
 B_{leak} & = & \frac{1}{4} B_{0}(\cos\theta_{++} - \cos\theta_{-+} - \cos\theta_{+-} + \cos\theta_{--}) \\
 B_{nr} & = & \frac{1}{4} B_{0}(\cos\theta_{++} + \cos\theta_{-+} - \cos\theta_{+-} - \cos\theta_{--})\ .
 \end{eqnarray}
\end{subequations}
Because of the possibility of both changing the relative angle and magnitude of the magnetic field upon reversal of either the applied electric or magnetic fields, there are many potential underlying causes for a non-zero derived value of a $B_{leak}$, $B_{corr}$, or $B_{nr}$ field parameter.  Note, however, that the simplest way outside of an electronics issue to generate a non-zero $B_{corr}$ would be to have misaligned fields.

\subsection{N-odd combinations}

There are four other combinations that can be formed from the measured frequencies.  These remaining combinations all contain the frequency difference between the $N=\pm1$ states, i.e., they are odd under reversal of $N$.  For a given state of the electric field and magnetic field, the difference of the precession frequencies for the two $N$ states will be given by equations similar to those for $\omega_{av,j,k}$ contained in Table~\ref{tab:meas_Efield}, except $2\bar{g}$ will be replaced by $\Delta g$.  In the limit where the molecule is fully polarized, which is very well satisfied under our conditions, $\Delta g/\bar{g} = k|\mathcal{E}|$ (see Eq.~\ref{eq:dgogvsE}).

Let us consider a simple model where the various states of the electric field are parametrized by three magnitudes.  The first parameter will be the average electric field magnitude $\mathcal{E}_{av}>0$.  If there is a fixed charge density on the walls of the vapor cell, the electric field may change its magnitude from $\mathcal{E}_{av}$ when the direction is reversed.  Let us denote this non-reversing component as $\mathcal{E}_{nr}$.  In addition to this non-reversing component of $\mathcal{E}$, let us consider the possibility that the electric field magnitude changes by an amount $\mathcal{E}_{corr}$ when the {\it absolute} magnetic field relative to the chosen laboratory frame is reversed\footnote{As with $B_{corr}$, one cause of $\mathcal{E}_{corr}$ could be unwanted electrical coupling between the magnetic and electric field power supplies.}.  A full table of this parametrization of the measured electric field in terms of these components for various states of the applied electric and magnetic field is shown in Table~\ref{tab:meas_Efield}.

These three parameters that describe the electric field combine with the four parameters to describe the magnetic field to become the seven field parameters used in our experiment.  Eight measured frequency combinations allow us to determine seven parameters describing experimental conditions, plus the value of the $e$EDM.  This parameterization therefore represents the maximum information that can be extracted from these frequency measurements alone.

Using this parameterization for the four states of the magnetic and electric fields, $\Delta_{-,+,+}\omega$ will be given by
\bn
 & \Delta_{-,+,+}\omega & \nonumber \\ 
 & = & k(\mathcal{E}_{av} + \mathcal{E}_{nr}+\mathcal{E}_{corr}) \bar{g} \mu_B(B_{av} + B_{corr} + B_{leak} + B_{nr}) \nonumber \\
  & & +k(\mathcal{E}_{av} - \mathcal{E}_{nr}+\mathcal{E}_{corr}) \bar{g}\mu_B(B_{av} - B_{corr} - B_{leak} + B_{nr}) \nonumber \\
  & & +k(\mathcal{E}_{av} + \mathcal{E}_{nr}-\mathcal{E}_{corr}) \bar{g}\mu_B(B_{av} + B_{corr} - B_{leak} - B_{nr}) \nonumber \\
  & & +k(\mathcal{E}_{av} - \mathcal{E}_{nr}-\mathcal{E}_{corr}) \bar{g}\mu_B(B_{av} - B_{corr} + B_{leak} - B_{nr}) \nonumber \\
  & = & 8 k \mathcal{E}_{av} \bar{g} \mu_B B_{av} + 8 k \mathcal{E}_{nr} \bar{g} \mu_B B_{corr} + 8 k \mathcal{E}_{corr} \bar{g} \mu_B B_{nr} \label{eq:D-++}\ .
\en
In the above equation, each line corresponds to the applied electric and magnetic fields being in a different state (see Table~\ref{tab:meas_Efield}).  In general, the misbehaving components of the magnetic and electric fields are small compared to the average fields, i.e., $|\mathcal{E}_{nr}|,|\mathcal{E}_{corr}|\ll\mathcal{E}_{av}$ and $|B_{nr}|,|B_{corr}|\ll B_{av}$.   Thus, Eq.~\ref{eq:D-++} should be dominated by $8 k \mathcal{E}_{av} \bar{g} \mu_B B_{av}$.

In a similar manner, $\Delta_{-,+,-}\omega$ may be expressed as
\bn
\Delta_{-,+,-}\omega & = & 8 k \mathcal{E}_{av} \bar{g} \mu_B B_{nr} + 8 k \mathcal{E}_{nr} \bar{g} \mu_B B_{leak} \nonumber \\
 & & \hspace{0.25in} + 8 k \mathcal{E}_{corr} \bar{g} \mu_B B_{av}\ ,
\en
and likewise $\Delta_{-,-,+}\omega$ may be expressed as
\bn
 \Delta_{-,-,+}\omega & = & 8 k \mathcal{E}_{av} \bar{g} \mu_B B_{corr} + 8 k \mathcal{E}_{nr} \bar{g} \mu_B B_{av} \nonumber \\
 & & \hspace{0.25in} + 8 k \mathcal{E}_{corr} \bar{g} \mu_B B_{leak}\ .
\en
This system of equations can be better expressed in terms of a matrix equation,
\bq
 \label{eq:matrix_for_Es}
 \frac{1}{2} k \left(\begin{array}{ccc}
  \Delta_{+,+,+}\omega & \Delta_{+,-,+}\omega & \Delta_{+,+,-}\omega \\ \Delta_{+,+,-}\omega & \Delta_{+,-,-}\omega & \Delta_{+,+,+}\omega \\ \Delta_{+,-,+}\omega & \Delta_{+,+,+}\omega & \Delta_{+,-,-}\omega \end{array}\right)\left(\begin{array}{c} \mathcal{E}_{av} \\ \mathcal{E}_{corr} \\ \mathcal{E}_{nr} \end{array}\right) = \left(\begin{array}{c} \Delta_{-,+,+}\omega \\ \Delta_{-,+,-}\omega \\ \Delta_{-,-,+}\omega \end{array}\right)
\eq
By solving this matrix equation, the electric field parameters can be determined from the measured frequency combinations.  % Using the work presented in Refs.~\cite{Lonseth1947,Dwyer1953}, the error in this inverse matrix may be determined and be used to determine the error in the parameters $\mathcal{E}_{av}$, $\mathcal{E}_{nr}$ and $\mathcal{E}_{corr}$.  In particular, the magnitude of the error of $\Delta_{+,+,+}$, $\Delta_{+,-,+}$, $\Delta_{+,+,-}$ and $\Delta_{+,-,-}$ will be the identical; therefore Eq.~(5.2) of Ref.~\cite{Dwyer1953} may be applied to determine the error bounds on the inverse matrix.

\begin{table}
 \center
  \begin{tabular}{lr}
  Label & Physical Quantity \\
  \hline
  $\Delta_{+,+,+}\omega$ & $16\bar{g}\mu_B B_{av}$ \\
  $\Delta_{-,+,+}\omega$ & $8 k \mathcal{E}_{av} \bar{g} \mu_B B_{av} + 8 k \mathcal{E}_{nr} \bar{g} \mu_B B_{corr} + 8 k \mathcal{E}_{corr} \bar{g} \mu_B B_{nr}$ \\
  $\Delta_{+,-,+}\omega$ & $16\bar{g}\mu_B B_{corr}$ \\
  $\Delta_{+,+,-}\omega$ & $16\bar{g} \mu_B B_{nr}$ \\
  $\Delta_{+,-,-}\omega$ & $16\bar{g} \mu_B B_{leak}$ \\
  $\Delta_{-,+,-}\omega$ & $8 k \mathcal{E}_{av} \bar{g} \mu_B B_{nr} + 8 k \mathcal{E}_{nr} \bar{g} \mu_B B_{leak} + 8 k \mathcal{E}_{corr} \bar{g} \mu_B B_{av}$ \\
  $\Delta_{-,-,+}\omega$ & $8 k \mathcal{E}_{av} \bar{g} \mu_B B_{corr} + 8 k \mathcal{E}_{nr} \bar{g} \mu_B B_{av} + 8 k \mathcal{E}_{corr} \bar{g} \mu_B B_{leak}$ \\
  $\Delta_{-,-,-}\omega$ & $8 k \mathcal{E}_{av} \bar{g} \mu_B B_{leak} + 8 k \mathcal{E}_{nr} \bar{g} \mu_B B_{nr} + 8 k \mathcal{E}_{corr} \bar{g} \mu_B B_{corr} - 16d_e\mathcal{E}_{eff}$ \\
  \hline\hline
 \end{tabular}
 \caption{\label{tab:field_reversals} Table of frequency combinations and associated field parameters, along with the $e$EDM value $d_e$.}
\end{table}

In addition to the electric field parameters, the $e$EDM also has an effect on the measured precession frequencies.  In particular, the $e$EDM causes the frequency difference between the $N=\pm1$ states to be shifted by $-4\text{sign}(\mathcal{E})\text{sign}(B)d_e\mathcal{E}_{eff}$.  Thus, for the above frequency combinations, the $e$EDM signal cancels.  But for the frequency combination that is odd in $N$, $\mathcal{E}$, and $B$, we find 
\bn
 \Delta_{-,-,-}\omega & = & 8 k \mathcal{E}_{av} \bar{g} \mu_B B_{leak} + 8 k \mathcal{E}_{nr} \bar{g} \mu_B B_{nr} + \nonumber \\
 \label{eq:edm_channel}
 & & \hspace{0.25in} 8 k \mathcal{E}_{corr} \bar{g} \mu_B B_{corr} - 16d_e\mathcal{E}_{eff}\ .
\en
A full listing of the frequency combinations is given in Table~\ref{tab:field_reversals}.  

In addition to the $e$EDM signal, we find three additional terms that can produce a non-zero $\Delta_{-,-,-}\omega$.  The middle two terms, $8 k \mathcal{E}_{nr} \bar{g} \mu_B B_{nr}$ and $8 k \mathcal{E}_{corr} \bar{g} \mu_B B_{corr}$, generally create small contributions to $\Delta_{-,-,-}\omega$, as they are products of a small component of the electric field and a small component of the magnetic field.  Of these three additional terms, the term $8 k \mathcal{E}_{av} \bar{g} \mu_B B_{leak}$ is generally anticipated to give the largest spurious contribution to $\Delta_{-,-,-}\omega$, as it depends on the product of a misbehaving component of the magnetic field and the average magnitude of the electric field.  However, compared to atomic experiments based on Eq.~\ref{eq:basic_larmor_precession} (i.e., without the internal comagnetometer feature that arises from the $\Omega$-doublet structure), the systematic error due to $B_{leak}$ is suppressed by $k\mathcal{E}_{av}\sim10^{-2}$.

\begin{figure}
 % Source of this figure is 2011-07-13/2054/step_0314.bin (chosen at random from the run).
 \center
 \includegraphics{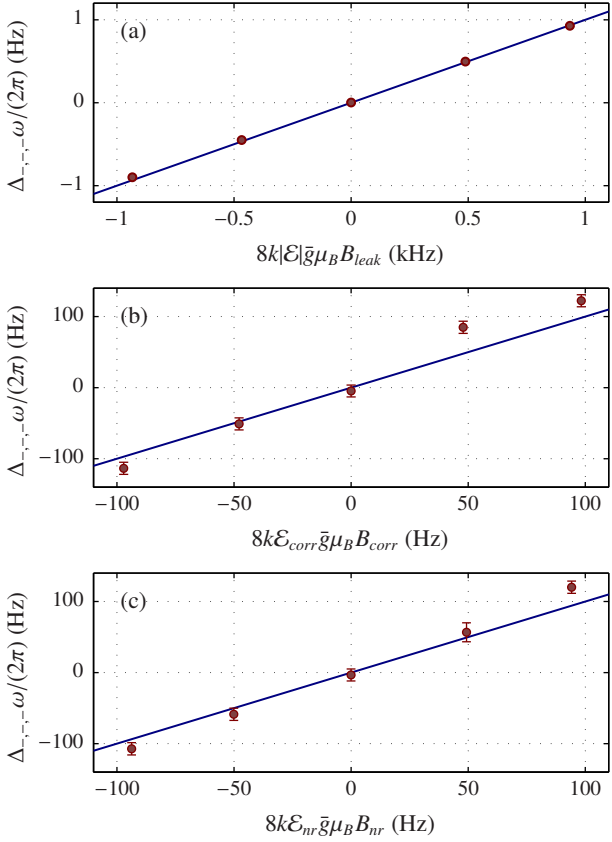}
 \caption{\label{fig:systematic_demonstration} (Color online) Shift in the $e$EDM channel, $\Delta_{-,-,-}\omega$ as a function of various products of field parameters, when such parameters are deliberately made large.  Each plot corresponds to a different product of misbehaving components of the electric and magnetic field.  The solid, blue lines correspond to a slope of one, which is the prediction of Eq.~\ref{eq:edm_channel} based on the units of the $x$ and $y$ axes, with no adjustable parameters.}
\end{figure}

To demonstrate this systematic rejection, we manually created large misbehaving components of the magnetic and electric fields and measured their impact on the $e$EDM signal channel.  The results of this study are shown in Fig.~\ref{fig:systematic_demonstration}.  The prediction of Eq.~\ref{eq:edm_channel} explains most of the correlation in the data ($\sim98\%$ for $B_{leak}$ and $\sim80\%$ for $B_{corr}$ and $B_{nr}$); however, the fits in Fig.~\ref{fig:systematic_demonstration} show reduced $\chi^2$ values significantly larger than unity.  The statistical uncertainties from the experiment are the only contribution used to compute these $\chi^2$ values, and therefore they might suggest suggest an additional systematic effect not accounted for in Eq.~\ref{eq:edm_channel}.

\section{The effect of field gradients}
\label{sec:gradients}
While the above discussion shows that there is significant power in the internal comagnetometer to determine the presence of systematic effects, the implicit assumption is that the applied fields are uniform and therefore identical for all molecules in the experiment.  Here we extend our discussion to address the following question: to what extent do magnetic and electric field gradients affect the field parameters of Sec.~\ref{sec:systematic_rejection} and $\Delta_{-,-,-}\omega$, the $e$EDM signal channel?

Perhaps the best proxy that exists for the size and strength of the field gradients in our experiment is the decay rate of the quantum beats $\Gamma=1/T_2^*$, as defined in Eqs.~\ref{eq:single_beat_fit}-\ref{eq:two_beat_fit}.  Two components contribute to the decay rate: the rate of dephasing of the beats due to all homogenous broadening effects $1/T_2$ (such as spontaneous emission and collisions) and the rate of dephasing of the beats due to field inhomogeneities $1/\tau$.  The former rate is strictly speaking unknown; however, we take the fluorescence decay rate of the $a$ state as a lower limit ($1/T_2\geq1/\tau_a$, where $\tau_a\approx50\us$; see Eq.~\ref{eq:background_fit})\footnote{The decay rate of the fluorescence itself has several components, e.g. the natural lifetime of the state and quenching of the electronic state due to collisions with the walls or other PbO, PbO$^*$, PbO$_2$, Pb$_2$O$_2$, etc. molecules.  However, it is possible that some collisions can cause dephasing, without quenching the $a$ state of the molecule.}.  For simplicity, we shall assume that $\Gamma = 1/T_2 + 1/\tau$.  With larger field gradients and therefore a larger distribution of precession frequencies, $\Gamma$ becomes larger.  If we assume that the additional rate of dephasing beyond $\Gamma = 1/T_2$ is due to these gradients, the fractional field inhomogeneity can then be estimated from the width of the peak in the the Fourier spectrum.  Namely, the width is given by 
\begin{equation}
 \label{eq:frac_inhomogeneity}
 \frac{\delta B}{B_{av}} = \frac{\delta \omega}{\omega_0} = \frac{2}{\tau \omega_0} = \frac{2}{\omega_0}\left(\Gamma-\frac{1}{T_2}\right)
\end{equation}
where $\delta B$ is the full-width, half maximum of the distribution of magnetic fields and $B_{av}$ is the average magnetic field, respectively, that is experienced by the ensemble of molecules.  Likewise, $\delta \omega = 2/\tau$ is the full-width, half maximum of the resulting distribution of precession frequencies and  $\omega_0$ is the resulting average precession frequency.

Equation~\ref{eq:frac_inhomogeneity} can be used to estimate an upper bound of the size of the gradients the ensemble of molecules experiences.  Without an electric field applied, only the absolute inhomogeneity of the $B$-field contributes to $\tau$, and a typical value of $\Gamma = 1/(37\us)$ for the beat decay rate at $\omega_0 = 2\pi\times(0.75\MHz)$ is observed.  Such a value, when combined with our upper limit of $T_2$, implies that $\delta B/B_{av}\leq0.29\%$.  At the much larger magnetic field $\omega_0 = 2\pi\times(2.25\MHz)$, the decay rate increases to $\Gamma = 1/(36\us)$, which actually implies a decrease in the upper limit of $\delta B/B_{av}$ to $\leq0.11\%$\footnote{These results imply that  $T_2$ is smaller than our upper bound, or the $B$-field inhomogeneity is not due primarily to the applied field, or some combination of the two.  If $T_2$ were equal to our upper bound and the field gradient was caused only by the applied magnetic field, then $\delta B/B_{av}$ would be constant.}.  When a maximally homogeneous electric field that is large enough to polarize the molecules is applied, the typical decay rate increases from $\Gamma\sim\!1/(35\us)$ to $\Gamma \sim\!1/(30\us)$ with an average $\omega_0 \approx 2\pi\times(3\MHz)$.  Because of Eq.~\ref{eq:larmorprec}, this implies that an application of an electric field increases the fractional inhomogeneity in the value of $\mathbf{B}\cdot\hat{\mathcal{E}}$ from $\lesssim\!0.09\%$ to $\lesssim0.13\%$\footnote{Here ``maximal homogeneity'' corresponds to conditions  where $\mathcal{E}<0$ and there is no electron emission as shown in Sec.~\ref{sec:apparatus}.  This configuration is discussed in more detail below.}.

\begin{figure}
 \center
 \includegraphics{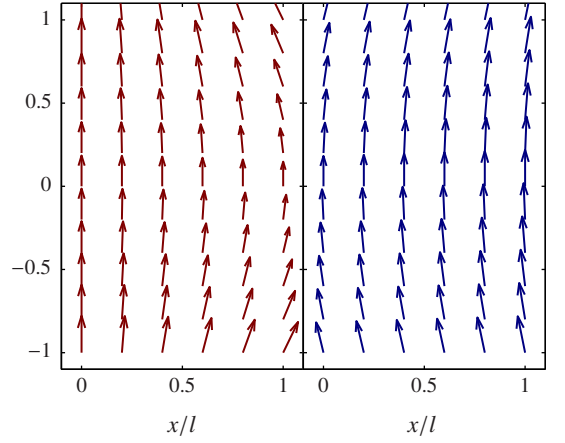}
 \caption{\label{fig:model_fields} Example electric (left) and magnetic field (right) profiles given by Eqs.~\ref{eq:example_efield} and ~\ref{eq:example_bfield}.  For the electric field, $a=0.2$ and for the magnetic field $b=0.2$.  Note that only the region $0<x<l$ is included because each detector in the experiment primarily captures fluorescence photons from one half of the cell.}
\end{figure}

Given that field gradients are observed in our experiment, the relevant question becomes to what extent do these gradients impact the measurement of the physical quantities listed in Table~\ref{tab:field_reversals}?  To answer this question, let us first consider an illuminating, yet simple model of a two-dimensional cross section of the gas of PbO, where electric field plates are positioned parallel to the $\hat{x}$ axis to generate an electric field in the $\hat{z}$ direction.  If the electric field plates are not infinite in extent, the electric field will experience fringing.  The effect of fringing will be to reduce the strength of the electric field as the edges of plates are approached.  Assuming that the reduction scales as $x^2$ to lowest order and the solution must obey Maxwell's equations, the resulting electric field profile will be given by
\bq
 \label{eq:example_efield}
 \bm{\mathcal{E}} = \mathcal{E}_0\left[1-a \left(\frac{x}{l}\right)^2 + a \left(\frac{z}{l}\right)^2\right]\hat{z} - 2 \mathcal{E}_0 a \frac{x z}{l^2}\hat{x}\ ,
\eq
where $\mathcal{E}_0$ is the strength of the electric field at $x=0$ and $z=0$, $l$ is the size of the area in which this approximation holds, and $a$ is a dimensionless parameter that describes the inhomogeneity of the field in the volume defined by $l$.  Because each detector in the experiment sees roughly half of the horizontal extent but the full vertical extent of the vapor cell, $x$ and $z$ must be constrained to $0<x<l$ and $-l<z<l$.  Let us also assume that there is an applied magnetic field in the $\hat{z}$ direction with a $dB_z/dx$ gradient.  To satisfy Maxwell's equations, the field must be given by
\bq
 \label{eq:example_bfield}
 \mathbf{B} = B_0\left(1 + b\frac{x}{l} \right)\hat{z} + B_0 b\frac{z}{l}\hat{x}\ ,
\eq
where $b$ describes the strength of the $dB_z/dx$ gradient and $B_0$ is the magnetic field at $x=0$ and $z=0$.  These two field profiles are shown in Fig.~\ref{fig:model_fields}.

Neglecting the contributions from components of the magnetic field perpendicular to the local direction of $\hat{\mathcal{E}}$ as well as the $e$EDM terms, Eq.~\ref{eq:larmorprec} yields simple expressions for the difference and average frequencies of the $N=\pm1$ states.  The measured difference frequency of the $N=\pm1$ states will be given by the average difference frequency over the volume, i.e., $\omega_d = \omega_+ - \omega_- = \langle 2 \bar{g} \mu_B (\mathbf{B}\cdot\hat{\mathcal{E}}) k \mathcal{E} \rangle = 2 k \bar{g} \mu_B \langle \mathbf{B}\cdot\bm{\mathcal{E}}\rangle$ where the brackets denote averaging over all space.  For the field gradients above, this averaging yields
\bn
 \omega_d & = & 2 k \bar{g} \mu_B \frac{1}{2l^2}\int_{0}^{l} dx \int_{-l}^{l} dz\ \mathbf{B}\cdot\bm{\mathcal{E}} \nonumber \\
 \label{eq:difffreq_gradient}
  & = & 2 k \bar{g} \mu_B B_0\mathcal{E}_0\left[1 + \left(\frac{1}{2} - \frac{5}{12} b\right) a\right]\ . 
\en
Similarly, the average frequency of the $N=\pm1$ states will be given by $\omega_{av} = \frac{1}{2}(\omega_+ + \omega_-) = 2 \bar{g} \mu_B \langle \mathbf{B}\cdot\hat{\mathcal{E}}\rangle$ or
\bq
 \omega_{av} = 2 \bar{g} \mu_B \frac{1}{2l^2}\int_{0}^{l} dx \int_{-l}^{l} dz\ \frac{\mathbf{B}\cdot\bm{\mathcal{E}}}{\mathcal{E}}\ .
\eq
The above integral can be approximated by expanding $1/\mathcal{E}$ in a power series in the gradient $a$ and then integrating.  The result of the integration is
\bq
 \label{eq:avfreq_gradient}
 \omega_{av} = 2 \bar{g} \mu_B B_0\left[1  + \frac{b}{2} - \frac{b a}{3} - \frac{2 (5 + 3 b) a^2}{45} + O(a^3)\right]
\eq
Thus we see that both the average frequency and the difference frequency can be affected by the presence of gradients.

The determination of the various electric field parameters of Sec.~\ref{sec:systematic_rejection} is predicated upon the relationship between the difference and average frequency, i.e., $\omega_{d} = k\mathcal{E}\omega_{av}$ or $\omega_+ - \omega_- = k\mathcal{E} \frac{1}{2}(\omega_+ + \omega_-)$ (see Eq.~\ref{eq:dgogvsE}).  Inserting Eqs.~\ref{eq:difffreq_gradient} and~\ref{eq:avfreq_gradient} into this relationship allows us to write an expression for the average derived electric field magnitude $\mathcal{E}_{der} = \omega_d/(k\omega_{av})$.  For our example field configuration, the derived electric field is given by
\bn
 \mathcal{E}_{der} & = & \mathcal{E}_0\left[1 - \frac{b a}{6 (2 + b)} \right. \nonumber \\ 
 & & \hspace{0.25in} \left. + \frac{(40 + 44 b + 7 b^2) a^2}{45 (2 + b)^2} + O(a^3) \label{eq:Ederived}\right]\ .
\en
Unlike what is assumed in Sec.~\ref{sec:systematic_rejection}, it is now apparent that the {\it electric field} magnitudes inferred from the data can depend on the strength of the background {\it magnetic field gradient} $b$.  Moreover, the accuracy of the average electric field magnitude measurement also depends on the electric field gradient $a$.  To demonstrate this, we calculate the actual average electric field magnitude,
\bn
 \langle \mathcal{E}\rangle & = & \frac{1}{2l^2}\int_{-l}^l\ dz\int_0^l\ dx \mathcal{E} \nonumber \\
  & \approx & \mathcal{E}_0\left[1 + \frac{2}{9}\mathcal{E}_0 a^2 - \frac{26}{525} a^4 + O(a^6)\right]\ ,
\en
and take its difference with the derived electric field\footnote{Note that for space considerations, the relevant $a^4$ term for this calculation is not shown in Eq.~\ref{eq:Ederived}.} when $b=0$,
\bq
 \mathcal{E}_{der} - \langle \mathcal{E}\rangle \approx \frac{704}{14175} \mathcal{E}_0 a^4 + O(a^6)\ .
\eq
Therefore, with this field configuration, the measurement of the magnitude of the electric field is impacted by the electric field gradients at fourth order.

If the magnetic and/or electric field gradients change when the fields are reversed, various frequency combinations $\Delta_{i,j,k}\omega$ can be non-zero, and therefore various field parameters such as $B_{corr}$, $\mathcal{E}_{corr}$ and $B_{leak}$ will be inferred to be non-zero as well.  In the context of our example, if we assume that the electric field gradients are unaffected by the magnetic field state and vice versa, we can define $a_{j}$ to be the electric field gradient parameter when the electric field has sign $j$.  Likewise, let us define $b_{k}$ to be the magnetic field gradient parameter when the magnetic field has sign $k$.  It can then be shown that for the field profile used above
\bn
 \mathcal{E}_{corr} & = & -\frac{\mathcal{E}_0}{12}\left[a_{+}(b_+ - b_-)+a_{-}(b_+ - b_-)\right] \label{eq:toymodel_Ecorr}\\
 B_{corr} & = & -\frac{B_0}{3}\left[(a_+-a_-)b_+ + (a_+-a_-)b_-\right] \\
 B_{leak} & = & -\frac{B_0}{3}\left[a_+b_+-a_-b_+ - a_+b_- + a_-b_-\right]\ .
\en
Note that if the magnetic field gradients are reversed perfectly, i.e., $b_{+}=b_{-}$, $\mathcal{E}_{corr}=0$.  Likewise, if the electric field gradients are reversed perfectly, then $a_-=a_+$ and $B_{corr}=0$.  To make $B_{leak} = 0$, {\it either} the electric field gradients or the magnetic field gradients must be reversed perfectly.  Therefore, to generate an apparent nonzero value of $B_{leak}$ through field gradients, both field gradients must change upon reversal.

One can imagine field profiles more complicated and more realistic than those used in the example above.  Such models could include misaligned fields (i.e. transverse field components), field gradients, and changing average magnitude of the fields.  For example, using a misaligned magnetic field together with Maxwell's conformal solution for the fringing electric field in a parallel-plate capacitor~\cite{Maxwell1904}, it can be shown that the derived value of the electric field magnitude will change as the magnetic field is rotated from the vertical direction.  The exact quantitative relationship between the magnetic and electric field parameters of Sec.~\ref{sec:systematic_rejection} and the field gradients depends strongly on the exact nature of the chosen field profile in any given model.  Because accurate field profiles cannot be determined with this apparatus, construction of an accurate quantitative model for this experiment is impossible.  Nevertheless, all of the models tested show the same qualitative behavior: changes in the magnetic field profile can affect the derived electric field parameters and vice versa.  To create a non-zero $\mathcal{E}_{corr}$ parameter, the magnetic field gradient must change when the magnetic field is reversed.  Likewise, to create a non-zero $B_{corr}$ parameter, the electric field profile must change when it is reversed.  Lastly, to create a non-zero $B_{leak}$ parameter or a false $e$EDM signal, both the electric and magnetic field profiles must change when the respective fields are reversed.

\begin{figure}
 \center
 \includegraphics{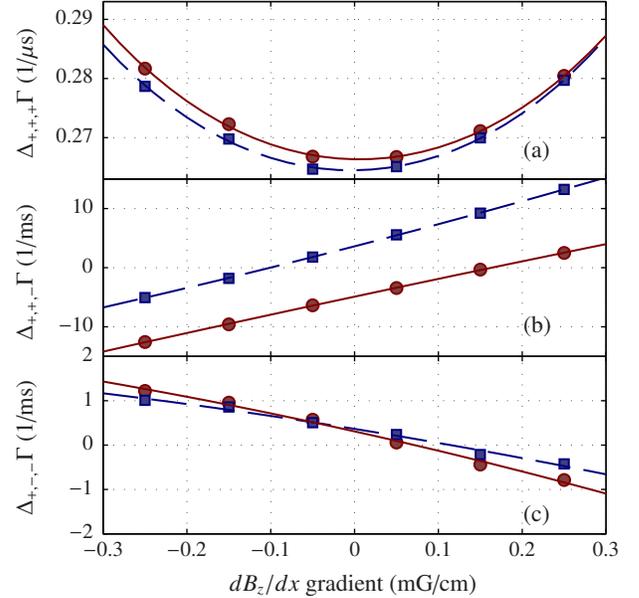}
 \caption{\label{fig:lifetime_comb} (Color online) Change in combinations of the beat decay rate as a function of an applied, fixed $dB_z/dx$ gradient.  The red circles show the signal from the detector fixed in the $+\hat{x}$ direction and the blue squares show the signal from the detector along the $-\hat{x}$ direction.  (a) The combination $\Delta_{+,+,+}\Gamma$ is eight times the decay rate averaged over all state and field configurations.  The combinations (b) $\Delta_{+,+,-}\Gamma$ and (c) $\Delta_{+,-,-}\Gamma$ show how the decay rate changes with reversal of the magnetic field and both the magnetic and electric fields, respectively.  The red, solid and blue, dashed lines are quadratic fits to the signals from the two detectors along $+\hat{x}$ and $-\hat{x}$, respectively.}
\end{figure}

While the situation may now appear intractable, it remains important to note that if the gradients change upon the reversal of the electric and/or magnetic fields, the beat decay rate $\Gamma$ must also change.  Therefore, forming combinations of $\Gamma$, analogous to the combinations of frequency discussed in Sec.~\ref{sec:systematic_rejection}, yields a quantitative measure of the amount of change in the field gradients.  For example, the combination $\Delta_{+,+,-}\Gamma$ quantifies how much the magnetic field gradient (or, rather, the inhomogeneity in $\mathbf{B}\cdot\hat{\mathcal{E}}$) changes when the magnetic field is reversed.

An example of this decay rate measurement and the subsequent combinations is shown in Fig.~\ref{fig:lifetime_comb}.  As a function of an applied, fixed $dB_z/dx$ gradient (henceforth, the term {\it fixed} shall refer to a gradient that does not reverse with its corresponding field, e.g., $b_+=-b_-$), the average decay rate $\Delta_{+,+,+}\Gamma$ changes and minimizes near zero gradient.  However, the difference in the decay rate when the magnetic field is reversed varies approximately linearly with an applied, fixed $dB_z/dx$ gradient, as shown in Fig.~\ref{fig:lifetime_comb}b.  Note that this difference in the decay rate when the magnetic field is reversed does not go to zero when the average decay rate $\Delta_{+,+,+}\Gamma$ is minimized, nor does it become zero for the two different detectors at the same applied $dB_z/dx$ gradient\footnote{This implies the presence of another magnetic field gradient in the system.}.  Lastly, a non-zero $\Delta_{+,-,-}\Gamma$ is also observed and is shown to have a dependence on an applied, fixed $dB_z/dx$ gradient.  This indicates a change in decay rate upon reversal of both the magnetic and electric fields.  

In principle, one can apply not only a fixed gradient, but also a gradient that reverses with the magnetic field (henceforth, the term {\it reversing} shall refer to a gradient that does reverse with its corresponding field, e.g., $b_+=b_-$).  A deliberately-applied, reversing component may be useful in order to make the applied magnetic field more uniform.  For example, if there is a displacement of the cell from the center of the Helmholtz coil, a linear field gradient that reverses with the applied field generated by the Helmholtz coil can shift the maximum of the field back to the center of the cell. Inside a magnetic shield, a Helmholtz coil will generate a field that has a quadratic gradient (e.g., a non-zero $d^2B_z/dx^2$).  If the geometric center of the cell and the center of the coil do not match, a linear gradient can be applied to effectively shift the maximum of the field to the center of the cell.  In doing so, such a linear gradient would have to be reversed with the field generated by the Helmholtz coils in order to keep the shifted maximum in the center of the cell.

\begin{figure}
 \center
 \includegraphics{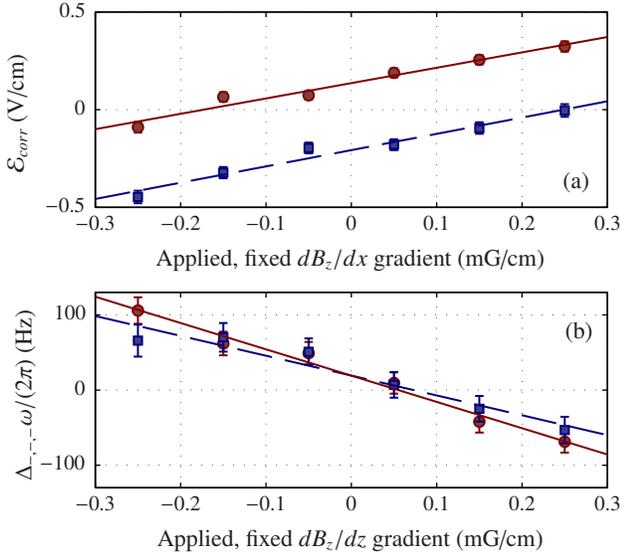}
 \caption{\label{fig:params_gradients} (Color online) (a) Variation of $\mathcal{E}_{corr}$ as a function of fixed $dB_z/dx$.  (b) The variation in $e$EDM signal channel $\Delta_{-,-,-}\omega$ as a function of fixed $dB_z/dz$.  The red circles show the signal from the detector fixed in the $+\hat{x}$ direction and the blue squares show the signal from the detector along the $-\hat{x}$ direction.}
\end{figure}

Because non-reversal of both electric and magnetic field gradients can shift the measured values of various parameters, a careful study was performed to determine the effects of various applied fixed and reversing magnetic field gradients on the field parameters of Sec.~\ref{sec:systematic_rejection}.  By deliberately applying both fixed and reversing components of the magnetic field gradients, we can quantitatively measure the effect of any particular gradient on $\Delta_{+,-,-}\Gamma$, $\mathcal{E}_{corr}$, $B_{leak}$, etc.  Shown in the top panel of Fig.~\ref{fig:params_gradients} is one example of the results if a magnetic field gradient is varied.  Here, a dependence of $\mathcal{E}_{corr}$ on a fixed $dB_z/dx$ gradient is observed.  Such a dependence of $\mathcal{E}_{corr}$ can be qualitatively explained using the toy model above, i.e., because the gradient does not reverse with the magnetic field, we set $b_{-} = -b_{+}$ and Eq.~\ref{eq:toymodel_Ecorr} becomes $\mathcal{E}_{corr}=\mathcal{E}_0(a_++a_-)b_+/6$.

\begin{figure}
 \center
 \includegraphics{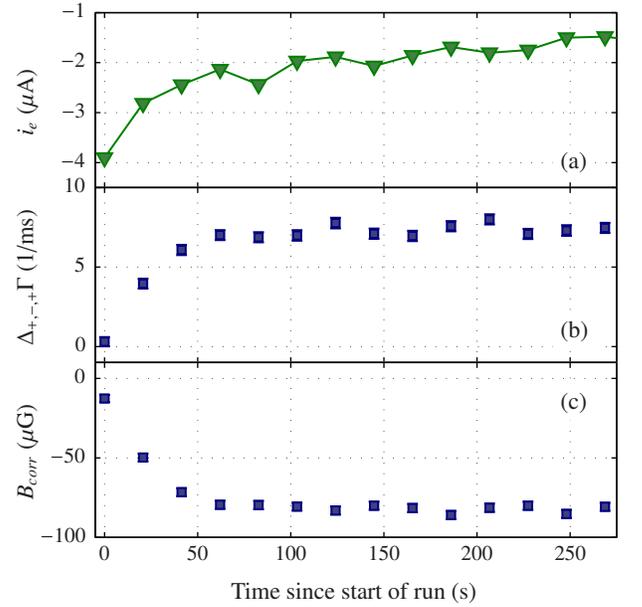}
 \caption{\label{fig:lifetime_time} (Color online) Change, as a function of time, in (a) the electron emission current, (b) $\Delta_{+,-,+}\Gamma$, and (c) $B_{corr}$,  when $\mathcal{E}>0$.  The electron emission current is defined as the sum of all of the currents flowing into or out of the cell through the electrodes and is assumed to behave in a similar way to Fig.~\ref{fig:electron_emission}.  At times $t<0$, $\mathcal{E}=0$.}
\end{figure}

A particularly interesting quantity is $\Delta_{+,-,+}\Gamma$, which quantifies how the gradients of the electric field change upon its reversal.  Thus, $\Delta_{+,-,+}\Gamma$ can show the effects and general behavior of electron emission described in Sec.~\ref{sec:apparatus}.  Consider the data shown in Fig.~\ref{fig:lifetime_time}.  At the start of the run, we observe a large current due to electron emission that becomes smaller with time but trends to a non-zero, steady-state value of approximately $1.5\uA$.  The time dependence of $\Delta_{+,-,+}\Gamma$ follows a similar function form, but starts at a zero value and trends to a non-zero, steady-state value.  This appearance of a non-zero $\Delta_{+,-,+}\Gamma$ indicates that there is a distortion in the electric field profile that is correlated with electron emission reaching its steady-state\footnote{While the causal relationship is not clear, it is possible that electron emission can lead to an equilibrium state where various insulators in the cell have trapped charges on their surfaces or unwanted voltage drops due to current flow.  Both of these effects could distort the electric field in such a way as to reduce the electric field near the emitting surface (thus reducing the emission current) but at the same time cause distortion of the electric field surrounding the molecules.}.  As the inhomogeneity of the electric field increases, a correlated change in $B_{corr}$ is observed.  In the context of the gradient example described above, such a $B_{corr}$ can arise if $a_+ \neq a_-$.  Such a $B_{corr}$ can also arise if electron emission leads to changes in the angle of $\bm{\mathcal{E}}$, causing $\theta_{+-}\neq\theta_{++}$ as described in Sec.~\ref{sec:systematic_rejection}.  Some combination of these scenarios must be expected due to the non-reversing nature of the electron emission current, combined with the voltage drops and associated $\bm{\mathcal{E}}$-fields that accompany this current.  

Compared to magnetic field gradients, the electric field gradients cannot be as well controlled in this apparatus.  However, it was empirically discovered that the size of the electric field gradients depends on the length of time the electric field is on during a data run ($T_{E\neq0}$) compared to the length of time when the electric field is not applied ($T_{E=0}$).  Reducing the duty cycle of the applied electric field, defined as $T_{E\neq0}/(T_{E=0}+T_{E\neq0})$, below 30\% reduces the size of $\Delta_{+,-,+}\Gamma$ and the misbehaving field parameters $\mathcal{E}_{nr}$, $B_{corr}$ and $B_{leak}$.

The effects of both fixed and reversing components of $dB_z/dz$, $dB_z/dx$, $dB_z/dy$, $dB_x/dx$, $dB_y/dy$, $B_x$ and $B_y$ on all the field parameters of Sec.~\ref{sec:systematic_rejection} and on the $e$EDM signal channel $\Delta_{-,-,-}\omega$ have been quantified.  Two of the 14 components mentioned are observed to impact $\Delta_{-,-,-}\omega$.  The first is a fixed, transverse field $B_x$, which our simple model above does not incorporate.  The second and only magnetic field gradient is $dB_z/dz$, as shown in Fig.~\ref{fig:params_gradients}b.  While the simple model above also does not incorporate a $dB_z/dz$ magnetic field gradient, the following linear gradients reproduce such an effect:
\bn
 \mathbf{B} & = & B_0\left(1 + b\frac{z}{l}\right)\hat{z} - b B_0 \frac{x}{l}\hat{x} \\
 \mathcal{E} & = & \mathcal{E}_0\left(1 + a\frac{z}{l}\right)\hat{z} - a \mathcal{E}_0\frac{x}{l}\hat{x}\ ,
\en
where the range of $x$ and $z$ is specified by $-l<x<l$ and $-l<z<l$, respectively.  With this configuration,
\bq
 \omega_{d} = 2\bar{g}k\mu_B\mathcal{E}_0B_0\left(1 + \frac{2}{3}ab\right)\ . 
\eq
Once again, if we assume the electric field profile is unaffected by the magnetic field and vice versa, then we can write $a_i$ ($b_i$) as the size of the electric (magnetic) field gradient when the field has sign $i$.  Then,
\bq
 \Delta_{-,-,-}\omega = \frac{4}{3}\bar{g}k\mu_B\mathcal{E}_0B_0(a_+b_+ - a_-b_+ - a_+b_- + a_-b_-)\ .
\eq
With a fixed $dB_z/dz$, $b_- = - b_+$ and
\bq
 \label{eq:d---f_fdBdz_theory}
 \Delta_{-,-,-}\omega = \frac{8}{3}\bar{g}k\mu_B\mathcal{E}_0B_0(a_+ - a_-)b_+\ .
\eq
Therefore, if the electric field gradient $d\mathcal{E}/dz$ does not reverse perfectly, it is not surprising that a fixed $dB_z/dz$ gradient should produce a false $e$EDM signal.  However, minimization of $dB_z/dz$ and other gradients can be achieved by minimizing the average beat decay rate $\Delta_{+,+,+}\Gamma$, as described in the the next section.

\section{EDM limit}
\label{sec:edm_limit}
Many types of experimental imperfections can conspire together to create non-zero values of the parameters describing the magnetic and electric fields, as enumerated in Sec.~\ref{sec:systematic_rejection}.  Parameters such as $B_{corr}$, $B_{nr}$ and $\mathcal{E}_{nr}$ can be generated by changing gradients, misalignment of the fields, non-reversing components of the magnetic or electric fields, or some combination thereof.  However, parameters such as $\mathcal{E}_{corr}$ should only be generated by inhomogeneities in the magnetic or electric field profile.  Given that a systematic $e$EDM signal can be generated by gradients (e.g., by a fixed $dB_z/dz$ gradient, as shown in Sec.~\ref{sec:gradients}), one must be careful to optimize the field profiles prior to taking any $e$EDM data.

\subsection{Minimizing gradients and optimizing the fields}
\label{sec:minimize_gradients}
In general, the gradients of the magnetic field tend to be minimized when the overall lifetime is maximized.  Using data similar to that shown in Fig.~\ref{fig:lifetime_comb}, we attempt to minimize the average beat decay rate $\Delta_{+,+,+}\Gamma$ using both fixed and reversing components of applied magnetic field gradients.  Given that we can apply both fixed and reversing components of five first-order gradients with our apparatus, this represents a difficult optimization problem for which a solution cannot always be found.  Moreover, the optimal solution may not make other combinations of the decay rate, such as $\Delta_{+,+,-}\Gamma$ or $\Delta_{+,-,+}\Gamma$, equal to zero.  Such a situation indicates that although the gradients are minimized, they may not be identical upon reversal of the magnetic or electric fields.

In order to gather useful $e$EDM data, the information gleaned above suggests a procedure for minimizing the spurious effects due to gradients.  By applying both fixed and reversing components of both $dB_z/dx$ and $dB_z/dy$, we can demonstrate conditions that minimize the gradient (minimize $\Delta_{+,+,+}\Gamma$) while equalizing the frequency measurements from the two detectors and driving $\Delta_{+,+,-}\Gamma \rightarrow 0$.  Moreover, minimizing $\Delta_{+,+,+}\Gamma$ with both fixed and reversing components of $dB_z/dz$ ensures that $dB_z/dz$ is minimized in both the positive and negative field configurations.  However, $\Delta_{+,+,+}\Gamma$ is close to minimized with no applied $dB_z/dz$ gradient; therefore, the applied $dB_z/dz$ was set to zero in the course of the $e$EDM data set.  These measures ensure the best magnetic field profile achievable.  To control electron emission and hence minimize its contributions to non-reversing electric field components, the temperature of the vapor cell is maintained near $665\degC$, the lowest temperature at which good signal to noise can be achieved.  In addition, the electric field duty cycle is set below 30\%, in order to ensure sufficient time for the non-Ohmic component of the leakage currents to settle to a small value.   These two measures help to control the change in the electric field correlated with electron emission.

Moreover, the derived values of the field parameters provide a measure of how well the fields and their respective gradients reverse.  In particular, a non-zero $\mathcal{E}_{corr}$ or $B_{nr}$ can indicate a magnetic field profile that changes upon reversal.  Likewise, a non-zero $\mathcal{E}_{nr}$ and $B_{corr}$ can indicate an electric field profile that changes upon reversal.  A non-zero $B_{leak}$ parameter could indicate the non-reversal of both magnetic and electric field profiles, for if even one reversed perfectly, this term would be zero.

\subsection{Data collection}
\label{sec:data_collection}
For a given data run, defined as data taken under the same experimental conditions, the data is collected according to a relatively standard procedure.  In order to calculate the field parameters of Sec.~\ref{sec:systematic_rejection} and the $e$EDM for a given data run, four applied field configurations must be used, corresponding to the four combinations of the signs of the magnetic and electric field.  In general, an applied field configuration is selected, and 512 laser shots (representing 5.12~s of data) are recorded with that field configuration.

For every laser shot, each quantum beat signal is fit to either Eq.~\ref{eq:single_beat_fit} or Eq.~\ref{eq:two_beat_fit}, depending on the data taking mode\footnote{For data taken with microwaves, the $N$ state selected for readout alternates from shot to shot, with the $N$ state probed during the first laser shot of 512 chosen at random.}.  The resulting collections of 512 best-fit values (e.g., $\Gamma$, $\omega_1$, $a_1$ and $\phi_1$) are then binned, with the number of bins determined by Scott's normal reference rule~\cite{Bevington2003}, and the resulting distribution fit to a Gaussian.  These distributions of best fit parameters show some outliers, which we believe are caused by fluctuations in the laser's output intensity\footnote{For large fluctuations of the laser intensity, often the size of the beat signal will differ significantly from the average signal of the full 512 laser shots.  The average signal is used to determine the initial guess of fit parameters (e.g., $\Gamma$, $a_1$, $\omega_1$) in the non-linear fitting algorithm.  For laser shots where the intensity drops significantly, the initial guess differs from the best fit parameters, that it is not guaranteed that the non-linear fitting algorithm will converge to the best solution.}.  For each fit parameter, outliers are determined by Chauvenet's criterion~\cite{Bevington2003} and these laser shots were excluded from the binning and averaging of all fit parameters.  For a collection of 512 shots, approximately 10 laser shots are typically excluded due to their best fit $a_1$ or $a_2$ values being significantly smaller than the mean.

After 512 laser shots are recorded, the field configuration is changed.  Within a given data set, each of the four field configurations is repeated approximately 16--32 times.  The particular temporal order of the four applied field configurations has been found to not affect the final result.  For example, with some runs, the electric field was reversed $\mathcal{N}$ times with the magnetic field positive or negative, the magnetic field was reversed and the magnetic shields degaussed, followed by another $\mathcal{N}$ reversals of the electric field.  For other runs, the four magnetic and electric field configurations were cycled sequentially, without degaussing of the magnetic fields.  No difference between the two sets of runs is observed.

With all field configurations measured $\mathcal{N}$ times, the combinations $\Delta_{i,j,k}\omega$ and $\Delta_{i,j,k}\Gamma$ are then computed as a function of time using the four closest-spaced collections of 512 laser shots with the four required field configurations.  Choosing the four collections most closely spaced in time minimizes the effects of long term drifts in the magnetic and electric fields.

\subsection{Statistics \& data constraints}

\begin{figure}
 \center
 \includegraphics{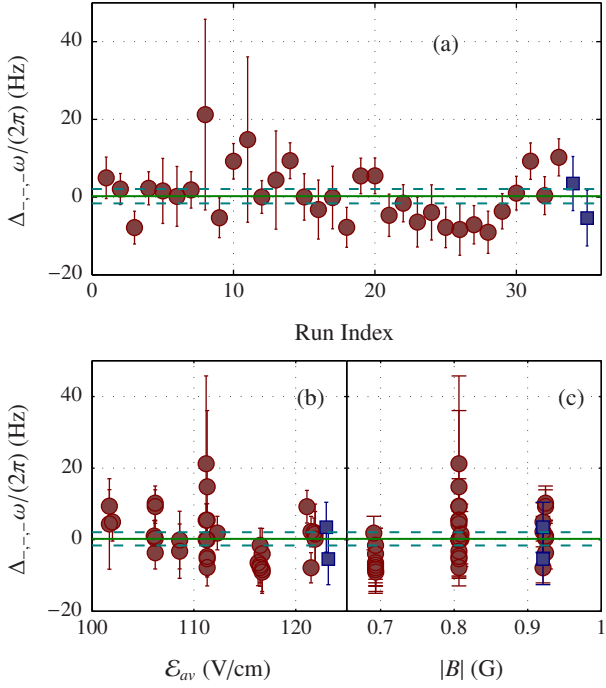}
 \caption{\label{fig:edm_statistics} (Color online) Final $e$EDM data set. (a) Best fit $e$EDM values for each data run contained in the final data set.  (b) The $e$EDM data shown as a function of electric field.  (c) The $e$EDM data shown as a function of magnetic field.  In each plot, the red circles represent data taken with two-beat technique, and the blue squares show data taken using microwave-erasure technique.  The green, solid line indicates the best average $\Delta_{-,-,-}\omega$, and the dashed, cyan lines indicate the 95\% confidence interval for that average value.}
\end{figure}

Using data taken under the optimal conditions as a guide, we constrain our full collection of data runs in an attempt to eliminate any of the spurious effects that were described in Sec.~\ref{sec:systematic_rejection}-\ref{sec:gradients}.  In order to ensure somewhat accurate reversal of the magnetic field gradients, we require that $|\Delta_{+,+,-}\Gamma/\Delta_{+,+,+}\Gamma|<0.0155$ and $|\mathcal{E}_{corr}/\mathcal{E}_{av}|<0.1\%$\footnote{These particular values were chosen to include the most data while minimizing the scatter in the field parameters and $\Delta_{-,-,-}\omega$.  For example, a change from $|\Delta_{+,+,-}\Gamma/\Delta_{+,+,+}\Gamma|<0.015$ to $0.016$ increases the $\chi_v^2$ of a Gaussian fit to the distribution of $\Delta_{-,-,-}\omega$ by almost a factor of 2.}.  To ensure somewhat accurate reversal of the electric fields, we require $|\mathcal{E}_{nr}/\mathcal{E}_{av}|<0.1\%$.

Using these constraints, a total of 4 hours of data is used in the final $e$EDM data set.  This data includes two-beat data in the range of $0.65\G\leq B_{av}\leq0.95\G$ (or, equivalently $1.7\MHz\leq 2\bar{g}\mu_BB_{av} \leq 2.4\MHz$) and $100\Vcm<\mathcal{E}_{av}<125\Vcm$.  It also includes microwave data taken with similar fields.  Lastly, some data where the procedure of Sec.~\ref{sec:minimize_gradients} was either not applied (26\% of the final data set) or partially applied (20\% of the final data set) manages to pass these cuts and therefore is included in the final data set.  The final set of data that passes all cuts is shown in Fig.~\ref{fig:edm_statistics}.

The $\Delta_{i,j,k}\omega$ and $\Delta_{i,j,k}\Gamma$ for each run are then averaged together, weighted by their respective errors.  The result for the $e$EDM channel is $\Delta_{-,-,-}\omega = 2\pi\times(0.20\pm0.91)\Hz$.  The $\chi_\nu^2$ value for the fit is 1.19 for 35 degrees of freedom; the probability for a larger $\chi_\nu^2$ to occur is approximately 21\%. %, which suggests that either the statistical errors are underestimated or there is an additional, unknown systematic effect that increases the size of the fluctuations within the data.  To compensate for this, we increase the size of the statistical error for our best fit $e$EDM by a factor of $\sqrt{\chi_\nu^2}=1.13$ to obtain the final statistical error of $\delta(\Delta_{-,-,-}\omega = 2\pi\times (\pm 1.04)$~Hz.

As a final note, there are no detectable differences between the microwave data and two-beat data.  However, the total amount of microwave data included is a factor of $\sim\!20$ below that of the two-beat data, making the statistical error approximately a factor $\sim\!4$ times larger.

\subsection{Systematic errors}
\label{sec:systematic_errors}

\begin{figure}
 \center
 \includegraphics{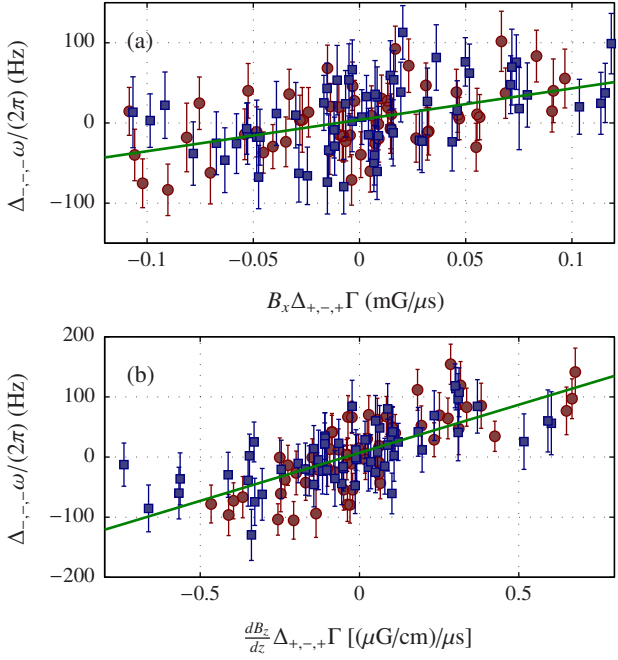}
 \caption{\label{fig:corr_example} (Color online) Observed correlation between $\Delta_{-,-,-}\omega$ and applied (a) $B_x$ and (b) $dB_z/dz$ while varying the electric field duty cycle, which impacts $\Delta_{+,-,+}\Gamma$ and therefore the reversibility of the electric field gradients.  The green lines represent the best linear least-squares fit of Eqs.~\ref{eq:corr_df_w_grad}-\ref{eq:corr_df_w_Bx}, and are used to extract the $c_{zz,EDM}$ and $c_{x,EDM}$ coefficients.  In both(a) and (b), the red circles show the signal from the detector fixed in the $+\hat{x}$ direction and the blue squares show the signal from the detector along the $-\hat{x}$ direction.}
\end{figure}

Even with the cuts described above, the data shows clear evidence of non-ideal electric and magnetic field reversals.  Therefore, an estimate of the error caused by the presence of imperfections of the magnetic and electric field profiles must be made.  There are two magnetic field imperfections, $dB_z/dz$ and $B_x$, that when coupled with electric field imperfections impact $\Delta_{-,-,-}\omega$ substantially.  In the case of $dB_z/dz$, Eq.~\ref{eq:d---f_fdBdz_theory} acts as a guide and suggests that, in the presence of a fixed magnetic field imperfection and an electric field gradient that changes with reversal of the electric field, the data should obey 
\bq
 \Delta_{-,-,-}\omega = c'_{zz,EDM} \frac{dB_z}{dz} \delta\mathcal{E} \label{eq:corr_df_w_grad_prime}\ ,
\eq
where $\delta\mathcal{E}$ is a measure of the changing electric field gradient and $c'_{zz,EDM}$ is the constant correlation coefficient.  To measure $c'_{zz,EDM}$, we deliberately apply large values of $dB_z/dz$ and increase the electric field duty cycle to amplify the negative effect of electron emission on the quality of the electric field.  Because there is no direct measurement of the imperfection of the electric field gradient reversal $\delta\mathcal{E}$, we use as a proxy the quantity $\Delta_{+,-,+}\Gamma$, which quantifies how much the beat decay rate changes when the electric field is reversed.  We use a simple first order approximation that the two are linearly proportional, i.e. $\Delta_{+,-,+}\Gamma \approx \zeta \delta\mathcal{E}$, where $\zeta$ is the first-order expansion constant.  Figure~\ref{fig:corr_example}b shows this data and the subsequent fit to   
\bq
 \Delta_{-,-,-}\omega = c_{zz,EDM} \frac{dB_z}{dz} \Delta_{+,-,+}\Gamma \label{eq:corr_df_w_grad}\ ,
\eq
where $c_{zz,EDM} = c'_{zz,EDM}/\zeta$ is the tunable constant extracted from the fit and $dB_z/dz$, $\Delta_{+,-,+}\Gamma$ and $\Delta_{-,-,-}\omega$ are the measured quantities.  The fit yields a value of $c_{zz,EDM}/(2\pi) = 160\pm15\mbox{ Hz$\cdot\mu$s/($\mu$G/cm)}$.  We note in passing that the simpler relation $\Delta_{-,-,-}\omega = \tilde{c}_{zz,EDM} \frac{dB_z}{dz}$, which ignores the impact of the changing electric field gradient, provides a poor fit to to data as $\frac{dB_z}{dz}$ is varied.

A similar dependence is observed in the case of $B_x$.  Empirically, the dependence of $\Delta_{-,-,-}\omega$ on $B_x$ is given by
\bq
  \Delta_{-,-,-}\omega = c_{x,EDM} B_x \Delta_{+,-,+}\Gamma \label{eq:corr_df_w_Bx}\ .
\eq
where $c_{x,EDM}$ is the correlation coefficient for a fixed $B_x$ component to the magnetic field.  This dependence is shown in Fig.~\ref{fig:corr_example}a.  The constant $c_{x,EDM}$ is measured in an analogous way to $c_{zz,EDM}$, and yields $c_{x,EDM}/(2\pi)=391\pm71\mbox{ Hz$\cdot\mu$s/mG}$.  

\begin{table*}
 \center
 \begin{tabular}{l|c|c|c|r|r|}
  Source & Formula & \multicolumn{2}{|c|}{Values} & Shift & Error \\
  \hline
  Fixed $dB_z/dz$ & $c_{zz,EDM}\frac{dB_z}{dz}\Delta_{+,-,+}\Gamma$ & $\frac{dB_z}{dz} = -7.6\pm8.1$~$\mu$G/cm  & $\Delta_{+,-,+}\Gamma=(0.23\pm0.20)$~1/ms & $-0.18$~Hz & $0.16$~Hz \\
  Fixed $B_x$ & $c_{x,EDM} B_x\Delta_{+,-,+}\Gamma$ & $B_x = -0.38\pm 0.16\mG$ & $\Delta_{+,-,+}\Gamma=(0.23\pm0.20)$~1/ms & $-0.034$~Hz & 0.030~Hz \\
  Product of $\mathcal{E}_{av}$ and $B_{leak}$  & $8 k \mathcal{E}_{av} \bar{g} \mu_B B_{leak}$ & $\mathcal{E}_{av}$ varies & $B_{leak}=-0.01\pm0.52\uG$ & $-0.004$~Hz & 0.048~Hz \\
  Product of $\mathcal{E}_{nr}$ and $B_{nr}$ & $8 k \mathcal{E}_{nr} \bar{g} \mu_B B_{nr}$ & $\mathcal{E}_{nr} = (51\pm9)\times10^{-3}\Vcm$ & $B_{nr}=5.7\pm8.0\uG$ & 0.00002~Hz & 0.00046~Hz \\
  Product of $\mathcal{E}_{corr}$ and $B_{corr}$ & $8 k \mathcal{E}_{corr} \bar{g} \mu_B B_{corr}$ & $\mathcal{E}_{corr}=(5\pm9)\times10^{-3}\Vcm $ & $B_{corr}=-2.5\pm2.4\uG$  & $-0.00014$~Hz & 0.00011~Hz \\
  \hline
  Total & & & & $-0.21$~Hz & 0.16~Hz \\
  \hline\hline
 \end{tabular}
 \caption{\label{tab:systematic_errors} Systematic error budget.  Shown are contributions  from each source considered in the analysis to $\Delta_{-,-,-}\omega/(2\pi)$, in units of Hz.  The sources shown are described in the text.}
\end{table*}

To estimate the systematic effect due to the $\frac{dB_z}{dz} \delta\mathcal{E}$ term on the $e$EDM data set, we first calculate a limit on the maximum size of any remnant $dB_z/dz$ during the $e$EDM measurement.  Two methods were used to measure this remnant $dB_z/dz$.  First, as described in Sec.~\ref{sec:minimize_gradients}, we attempted to cancel any fixed, remnant $dB_z/dz$ by applying a $dB_z/dz$ gradient to cancel the remnant gradient and therefore minimize $\Delta_{+,+,+}\Gamma$.  This procedure found that the background $dB_z/dz$ was close to zero and therefore we applied no $dB_z/dz$ gradient.  Over the course of the $e$EDM data set, this measurement was repeated at various stages and all measurements are included in computing the final, average value of $dB_z/dz$.  The second method involves masking the excitation laser such that only molecules on the top half of the vapor cell are excited into the $a$ state.  On the subsequent 512 laser shots, the mask is switched such that only molecules on the bottom half of the vapor cell are excited into the $a$ state.  The applied $dB_z/dz$ that makes the difference in the measured frequency for molecules on the top and bottom halves of the cell zero is the applied $dB_z/dz$ that cancels the background magnetic field gradient.  All these measurements are combined and yield an average fixed, remnant value of $dB_z/dz = -7.6\pm8.1$~$\mu$G/cm. %We therefore take as a conservative limit that $dB_z/dz<18$~$\mu$G/cm at 90\% confidence for the $e$EDM data. 

An estimate of the background, fixed component of $B_x$ must also be obtained in order to calculate the systematic effect of this transverse field.  Such a limit is taken from measurements of the misalignment of the electric and magnetic fields (see Sec.~\ref{sec:transverse_fields}).  These measurements suggest that the background, fixed $B_x$ field component is $B_x = -0.38\pm 0.16\mG$.  Because this value for the background, fixed $B_x$ is consistent with zero at the $3\sigma$ level, no compensation was applied to eliminate it.%We conservatively take $|B_x|<0.6\mG$ at 90\% confidence for the $e$EDM data set.

For each run indexed by $k$, the measured value of $\Delta_{+,-,+}\Gamma$ and the estimates of $dB_z/dz$ and $B_x$ are used to determine their effect on the $e$EDM channel for that run, $(\delta\Delta_{-,-,-}\omega)_k$ (using Eqs.~\ref{eq:corr_df_w_grad}-\ref{eq:corr_df_w_Bx}).  To determine the mean effect of the systematic on the full data set, the $(\delta\Delta_{-,-,-}\omega)_k$ are averaged together according to
\bq
 \label{eq:final_sys_shift}
 \overline{\delta\Delta_{-,-,-}\omega} = \left(\sum_k \frac{(\delta\Delta_{-,-,-}\omega)_k}{\sigma_{\Delta_{-,-,-}\omega,k}^2} \right)/\left(\sum_k \frac{1}{\sigma_{\Delta_{-,-,-}\omega,k}^2} \right)\ ,
\eq
where $\sigma_{\Delta_{-,-,-}\omega,k}$ is the statistical uncertainty in the $e$EDM value for run $k$.  Note that the final systematic error is determined by weighting the magnitude of the individual systematic errors by the statistical error for that run.  Each value of $(\delta\Delta_{-,-,-}\omega)_k$ has a corresponding uncertainty $\sigma_{\delta\Delta_{-,-,-}\omega,k}$, and using simple error propagation, we compute the resulting uncertainty in $\overline{\delta\Delta_{-,-,-}\omega}$ as
\bq
 \label{eq:final_sys_error}
 \sigma^2_{\overline{\delta\Delta_{-,-,-}\omega}} = \left(\sum_k \frac{\sigma^2_{\delta\Delta_{-,-,-}\omega,k}}{\sigma_{\Delta_{-,-,-}\omega,k}^2} \right)/\left(\sum_k \frac{1}{\sigma_{\Delta_{-,-,-}\omega,k}^2} \right)\ .
\eq
Alternatively, the standard deviation of the weighted average can also be computed.  Such a standard deviation can be used to estimate the uncertainty in the mean value.  If there are $\mathcal{N}$ runs, this estimate of the uncertainty of the mean systematic effect would be given by
\bq
 \label{eq:final_sys_error2}
 \sigma'^2_{\overline{\delta\Delta_{-,-,-}\omega}} = \frac{1}{\mathcal{N}}\left(\sum_k \frac{[(\delta\Delta_{-,-,-}\omega)_k - \overline{\delta\Delta_{-,-,-}\omega}]^2}{\sigma_{\Delta_{-,-,-}\omega,k}^2} \right)/\left(\sum_k \frac{1}{\sigma_{\Delta_{-,-,-}\omega,k}^2} \right)\ .
\eq
For any given systematic effect, we take the larger of $\sigma'_{\overline{\delta\Delta_{-,-,-}\omega}}$ or $\sigma_{\overline{\delta\Delta_{-,-,-}\omega}}$ as the final systematic uncertainty.

A breakdown of the most important individual contributions to the systematic error is shown in Table~\ref{tab:systematic_errors}.  For errors caused by magnetic field imperfections, we explicitly include contributions only from the two largest observed sources of correlation with the $e$EDM channel, namely those due to a fixed $dB_z/dz$ gradient or a fixed $B_x$ field.  Once again, our analysis indicates that these enter the $e$EDM channel due to their coupling to uncontrolled, non-reversing electric field gradients.  For reference, Table~\ref{tab:systematic_errors} also contains the systematic contributions to $\Delta_{-,-,-}\omega$ described in Sec.~\ref{sec:systematic_rejection}.  These contributions are calculated directly from the data, namely by constructing all possible combinations $\Delta_{i,j,k}\omega$.  The corresponding physical field parameters for each run, including $B_{leak}$, $B_{nr}$, $\mathcal{E}_{corr}$, etc., are computed from these combinations via Table~\ref{tab:field_reversals} and Eq.~\ref{eq:matrix_for_Es}.  With the field parameters, the systematic error for each run is computed using Eq.~\ref{eq:edm_channel}.  Eqs.~\ref{eq:final_sys_shift}-\ref{eq:final_sys_error2} are used to average together all the field parameters and systematic errors for the full data set, and the results are shown in Table~\ref{tab:systematic_errors}.

It is important to note that the largest systematic error from the type of uniform field parameters discussed in Sec.~\ref{sec:systematic_rejection} is $8\bar{g}\mu_B k\mathcal{E}_{av}B_{leak}$.  While $B_{leak}$ can be created by changing magnetic and electric field gradients, it can also be generated simply by a leakage current.  Given that the measured leakage currents in the cell are of the order of 10~$\mu$A, it is certainly plausible that these leakage currents could generate a $B_{leak}$ within our error of $0.53\uG$.  However, if this experiment was performed with a system with no internal comagnetometer, i.e.,  where the Larmor precession frequency is given by Eq.~\ref{eq:basic_larmor_precession}, the contribution to the uncertainty of the systematic shift from such a large leakage current would be a factor of $1/(k\mathcal{E}_{av})\sim100$ larger than it is here.  Hence, without the internal comagnetometer, the leakage current contribution would be the largest single contributor to the systematic error.

Finally, we note in passing that we also considered other possible sources of systematic error.  One example is possible errors arising from differences between the density and/or velocity of the populations in the $N=\pm1$ states.  By changing the detuning of the laser from the center of the Doppler-broadened line, we could selectively populate one $N$ state more than another.  By removing the retro-reflecting mirror and setting the laser detuning in the middle of the Doppler broadened line, each $N$ state would correspond to an equal and opposite velocity class, which can lead to a spatial separation of the two populations.  We found no dependence of $\Delta_{-,-,-}\omega$ on these effects.

We therefore quote the final values of $\Delta_{-,-,-}\omega = 0.20+0.22_\text{syst}\pm 0.91_\text{stat}\pm 0.17_\text{syst})$~Hz and $d_e = (-4.4\pm9.5_\text{stat}\pm1.8_\text{syst})\times10^{-27}\ecm$, where stat denotes the statistical 1$\sigma$ error and syst denotes the systematic shift and its corresponding 1$\sigma$ error.  A limit on the magnitude of the $e$EDM is obtained by integrating the assumed underlying  Gaussian distribution symmetrically about the mean value, with the standard deviation taken as the quadrature sum of the statistical and systematic errors.  The result is $|d_e|<1.7\times10^{-26}\ecm$, at 90\% confidence.

\section{Conclusion}
We have demonstrated that the $a^3\Sigma^+(v=5,J=1)$ $\Omega$-doublet state of PbO is capable of good systematic rejection and control in an electron electric dipole moment search.  Using PbO, we have obtained a limit of $|d_e|<1.7\times10^{-26}\ecm$ (90\% confidence), only about a factor of 20 worse than the world's best experimental limit~\cite{Hudson2011}.  Given the presence of significant non-reversing electric and magnetic field gradients (the former apparently due to uncontrolled electron emission from the top electrode) and leakage currents on the order of $10\uA$, obtaining such a limit provides a clear example of the power and flexibility an $\Omega$-doublet state gives toward diagnosing and controlling systematic errors in this type of experiment.

The primary reason for the demonstrated level of systematic rejection stems from the use of the two $N$ states that allow for measurement of the average magnetic field in all configurations of the experiment.  For this reason, this $\Omega$-doublet structure has been referred to as an ``internal comagnetometer''.  Compared to a traditional comagnetometer~\cite{KhriplovichLamoreaux1997}, this system is sensitive to exactly the same magnetic field with which the molecules used to detect the $e$EDM are interacting.  Moreover, the ability to accurately measure the electric field using the molecules, because of the dependence of the $g$ factor on electric field, gives even more information than a traditional comagnetometer.

Perhaps the only difficulty with this type of level structure is the breakdown of the comagnetometer function in the presence of magnetic and electric field gradients, rather than simply uniform fields.  This breakdown is caused by the ensemble of molecules being distributed in a finite volume with non-uniform magnetic and electric fields.  This complication can have an impact on the inferred field parameters such as the average electric field and the component of the magnetic field that mimics a field due to leakage currents.  However, by utilizing all available information, such as changes in the quantum beat decay rate and the derived electric and magnetic field parameters, large effects on the $e$EDM signal can be avoided.

The next generation of experiments that use molecules with similar level structure, such as those based on ThO~\cite{Vutha2010}, HfF$^+$\cite{Loh2011,Grau2012} and WC~\cite{Lee2009}, should, by extension, have similarly good systematic rejection.  In these experiments, the breakdown of the magnetometer due to gradients should pose less of a problem, as none require the use of a high-temperature vapor cell and thus avoid the complications inherent to the experimental apparatus described here.  For example, in the ongoing ThO experiment~\cite{Vutha2011}, which uses a molecular beam, the electric and magnetic fields are much more uniform than in this work.  Moreover, the fields are more well understood in that experiment; therefore more realistic and detailed modeling of the shifts encountered due to gradients can be undertaken.  Given the result presented in this paper, we expect that these future experiments should obtain dramatically better systematic rejection than that obtained here.

\begin{acknowledgments}
The authors thank S.K. Lamoreaux and A.O. Sushkov for useful discussions; S. Bickman, Y. Jiang, F. Bay, and D. Kawall for earlier contributions to the development of the experiment, and B. O'Leary for useful discussions related to the frequency combinations.  This work was supported by the NSF.
\end{acknowledgments}

\appendix

\section{Systematic rejection when measuring phase}
\label{app:phase_comb}

\begin{figure}
 \center
 \includegraphics{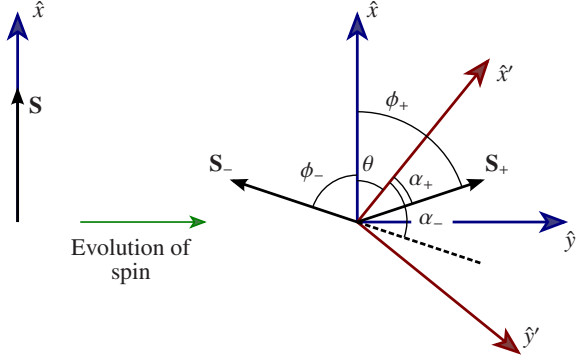}
 \caption{\label{fig:detection_geometry} Detection geometry for an $e$EDM experiment that measures phase.  See Appendix~\ref{app:phase_comb} for description. }
\end{figure}

\begin{table*}
 \center
 \begin{tabular}{llr}
   Type & Label & Physical Quantity \\
   \hline
   Phase & $\Delta_{+,+,+}\alpha$ & $16\bar{g}\mu_B B_{av}\tau$ \\
   & $\Delta_{-,+,+}\alpha$ & $8 k \mathcal{E}_{av} \bar{g} \mu_B B_{av}\tau + 8 k \mathcal{E}_{nr} \bar{g} \mu_B B_{corr}\tau + 8 k \mathcal{E}_{corr} \bar{g} \mu_B B_{nr}\tau$ \\
  & $\Delta_{+,-,+}\alpha$ & $16\bar{g}\mu_B B_{corr}\tau$ \\
  & $\Delta_{+,+,-}\alpha$ & $16\bar{g} \mu_B B_{nr}\tau - 8\theta$ \\
  & $\Delta_{+,-,-}\alpha$ & $16\bar{g} \mu_B B_{leak}\tau$ \\
  & $\Delta_{-,+,-}\alpha$ & $8 k \mathcal{E}_{av} \bar{g} \mu_B B_{nr}\tau + 8 k \mathcal{E}_{nr} \bar{g} \mu_B B_{leak}\tau + 8 k \mathcal{E}_{corr} \bar{g} \mu_B B_{av}\tau$ \\
  & $\Delta_{-,-,+}\alpha$ & $8 k \mathcal{E}_{av} \bar{g} \mu_B B_{corr}\tau + 8 k \mathcal{E}_{nr} \bar{g} \mu_B B_{av}\tau + 8 k \mathcal{E}_{corr} \bar{g} \mu_B B_{leak}\tau$ \\
  & $\Delta_{-,-,-}\alpha$ &  $8 k \mathcal{E}_{av} \bar{g} \mu_B B_{leak}\tau + 8 k \mathcal{E}_{nr} \bar{g} \mu_B B_{nr}\tau + 8 k \mathcal{E}_{corr} \bar{g} \mu_B B_{corr}\tau - 16d_e\mathcal{E}_{eff}\tau$ \\
  \hline
    Asymmetry & $\Delta_{+,+,+}$ & $-32\bar{g}\mu_B\delta B\tau$ \\
   & $\Delta_{-,+,+}A$ & $-16 k \mathcal{E}_{av} \bar{g} \mu_B B_{av}\tau - 16 k \mathcal{E}_{nr} \bar{g} \mu_B B_{corr}\tau - 16 k \mathcal{E}_{corr} \bar{g} \mu_B B_{nr}\tau$ \\
  & $\Delta_{+,-,+}A$ & $-32\bar{g}\mu_B B_{corr}\tau$ \\
  & $\Delta_{+,+,-}A$ & $-32\bar{g} \mu_B B_{nr}\tau + 16\theta$ \\
  & $\Delta_{+,-,-}A$ & $-32\bar{g} \mu_B B_{leak}\tau$ \\
  & $\Delta_{-,+,-}A$ & $-16 k \mathcal{E}_{av} \bar{g} \mu_B B_{nr}\tau - 16 k \mathcal{E}_{nr} \bar{g} \mu_B B_{leak}\tau - 16 k \mathcal{E}_{corr} \bar{g} \mu_B B_{av}\tau$ \\
  & $\Delta_{-,-,+}A$ & $-16 k \mathcal{E}_{av} \bar{g} \mu_B B_{corr}\tau - 16 k \mathcal{E}_{nr} \bar{g} \mu_B B_{av}\tau - 16 k \mathcal{E}_{corr} \bar{g} \mu_B B_{leak}\tau$ \\
  & $\Delta_{-,-,-}A$ & $-16 k \mathcal{E}_{av} \bar{g} \mu_B B_{leak}\tau - 16 k \mathcal{E}_{nr} \bar{g} \mu_B B_{nr}\tau - 16 k \mathcal{E}_{corr} \bar{g} \mu_B B_{corr}\tau + 32d_e\mathcal{E}_{eff}\tau$\\
  \hline\hline
 \end{tabular}
 \caption{\label{tab:fullcomb} Full phase and asymmetry combinations for an $e$EDM experiment that measures total precession through a magnetic field.}
\end{table*}

In case of an experiment detecting phase precession in a beam (e.g., the ThO $e$EDM experiment~\cite{Vutha2010}), most of the discussion of Sec.~\ref{sec:systematic_rejection} still applies; however, special attention must be dedicated to how the phase angle is detected.  In some experiments, it is possible to determine not only the total amount of phase precession, but also the direction of the rotation.  This is the case in Ramsey's method of separated oscillatory fields (SOF) when one uses rotating fields at the start and end of the sequence.  For this discussion, however, consider the case where the direction of spins' precession is detected by rotating the detection coordinate system relative to the preparation coordinate system by an angle $\theta$.  This is possible in the context of Ramsey's method of SOF if the two oscillating fields have different phases.

In particular, consider an experiment where one detects the phase angle in a coordinate system defined by the axes $\hat{x}'$ and $\hat{y}'$ that is rotated by an angle $\theta$ with respect to the coordinate frame where the $\hat{x}$ coordinate is defined by the initial orientation of the spin, as shown in Fig.~\ref{fig:detection_geometry}.  After evolving for a time $\tau$, the spin will have precessed by an angle $\phi_+$ if the spin is rotating in the clockwise direction.  Let us assume that clockwise rotation corresponds to $B>0$.  If the $B$ field is negative, the spin will therefore rotate through an angle $\phi_- = -\phi_+$.  After precessing for a time $\tau$ in a negative (positive) $B$ field, the measured angle in the detection coordinate system will be $\alpha_-$ ($\alpha_+$).  In this case, it is straightforward to show from Fig.~\ref{fig:detection_geometry} and the parameterization used in Table~\ref{tab:meas_Efield} that the various $\alpha_{av,j,k}$ are given by
\bn
 \alpha_{av,+,+} & = & 2\bar{g}\mu_B(B_{av} + B_{corr} + B_{leak} + B_{nr})\tau - \theta \\
 \alpha_{av,-,+} & = & 2\bar{g}\mu_B(B_{av} - B_{corr} - B_{leak} + B_{nr})\tau - \theta \\
 \alpha_{av,+,-} & = & 2\bar{g}\mu_B(B_{av} + B_{corr} - B_{leak} - B_{nr})\tau + \theta \\
 \alpha_{av,-,-} & = & 2\bar{g}\mu_B(B_{av} - B_{corr} + B_{leak} - B_{nr})\tau + \theta \ .
\en
The $N$-even combinations then become
\bn
 \Delta_{+,+,+}\alpha & = & 16\bar{g}\mu_BB_{av}\tau \\
 \Delta_{+,-,+}\alpha & = & 16\bar{g}\mu_B B_{corr}\tau \\
 \Delta_{+,+,-}\alpha & = & 16\bar{g}\mu_B B_{nr}\tau -8\theta  \\
 \Delta_{+,-,-}\alpha & = & 16\bar{g}\mu_B B_{leak}\tau\ .
\en
Note that adding an offset in the detection coordinates looks as if one is inducing a non-reversing component of the magnetic field.

One possible way to detect the angle $\alpha$ involves projecting the spin along either the $\hat{x}'$ axis (with probability $\cos^2\alpha$) and the $\hat{y}'$ axis (with probability $\sin^2\alpha$)~\cite{Vutha2010} .  One then defines the asymmetry
\bq
 A = \frac{N_x - N_y}{N_x + N_y} = \frac{\cos^2\alpha - \sin^2\alpha}{\cos^2\alpha + \sin^2\alpha} = \cos(2\alpha)\ .
\eq
Strictly speaking the $A$'s are not additive and thus cannot be used to compute the combinations except in an approximation where $\alpha \approx \pi/4$.  In that case, 
\bq
 A \approx -2\left(\alpha-\frac{\pi}{4}\right) \, 
\eq
and the average asymmetries become
\bn
 A_{av,+,+} & = & -4\bar{g}\mu_B(B_{av} + B_{corr} + B_{leak} + B_{nr})\tau + 2\theta + \frac{\pi}{2} \hspace{0.25in} \\
 A_{av,-,+} & = & -4\bar{g}\mu_B(B_{av} - B_{corr} - B_{leak} + B_{nr})\tau + 2\theta + \frac{\pi}{2} \hspace{0.25in} \\
 A_{av,+,-} & = & -4\bar{g}\mu_B(B_{av} + B_{corr} - B_{leak} - B_{nr})\tau - 2\theta + \frac{\pi}{2} \hspace{0.25in} \\
 A_{av,-,-} & = & -4\bar{g}\mu_B(B_{av} - B_{corr} + B_{leak} - B_{nr})\tau - 2\theta + \frac{\pi}{2} \hspace{0.25in} \ .
\en
Using the above expressions to form the $N$-even combinations yields the following result:
\bn
 \Delta_{+,+,+}A & = & -32\bar{g}\mu_B B_{av}\tau + 4\pi\\
 \Delta_{+,-,+}A & = & -32\bar{g}\mu_B B_{corr}\tau \\
 \Delta_{+,+,-}A & = & -32\bar{g}\mu_B B_{nr}\tau + 16\theta \\
 \Delta_{+,-,-}A & = & -32\bar{g}\mu_B B_{leak}\tau \ .
\en
One can then define $\delta B =   B_{av} - (\pi/4)/(2\bar{g}\mu_B\tau)$, which is the magnetic field magnitude that moves the average spin precession away from $\alpha=\pi/4$.  With this definition, 
\bq
  \Delta_{+,+,+}A = -32\bar{g}\mu_B \delta B \tau\ ,
\eq
which more closely resembles the expressions for $\Delta_{+,+,+}\omega$ and $\Delta_{+,+,+}\alpha$.

For the $N$-odd combinations, the resulting expressions are almost identical to the case of measuring frequency.  Consider first the case of detecting the asymmetry $A$; both the $\theta$ and $\pi/2$ terms cancel in the $N$-odd combinations, leaving expressions identical to those for $\Delta_{-,j,k}\omega$ except they are multiplied by $-2\tau$    Likewise, in the case of detecting the phase $\alpha$, the constant $\theta$ terms cancel in the $N$-odd combinations.  Thus, the expressions for $\Delta_{-,j,k}\alpha$ are  identical to those for $\Delta_{-,j,k}\omega$ except they are multiplied by $\tau$.  All of these results are enumerated for reference in Table~\ref{tab:fullcomb}.


\begin{thebibliography}{10}%
\makeatletter
\providecommand \@ifxundefined [1]{%
 \ifx #1\undefined \expandafter \@firstoftwo
 \else \expandafter \@secondoftwo
\fi
}%
\providecommand \@ifnum [1]{%
 \ifnum #1\expandafter \@firstoftwo
 \else \expandafter \@secondoftwo
\fi
}%
\providecommand \enquote [1]{``#1''}%
\providecommand \bibnamefont  [1]{#1}%
\providecommand \bibfnamefont [1]{#1}%
\providecommand \citenamefont [1]{#1}%
\providecommand\href[0]{\@sanitize\@href}%
\providecommand\@href[1]{\endgroup\@@startlink{#1}\endgroup\@@href}%
\providecommand\@@href[1]{#1\@@endlink}%
\providecommand \@sanitize [0]{\begingroup\catcode`\&12\catcode`\#12\relax}%
\@ifxundefined \pdfoutput {\@firstoftwo}{%
 \@ifnum{\z@=\pdfoutput}{\@firstoftwo}{\@secondoftwo}%
}{%
 \providecommand\@@startlink[1]{\leavevmode}%
 \providecommand\@@endlink[0]{}%
}{%
 \providecommand\@@startlink[1]{%
  \leavevmode
  \pdfstartlink
   attr{/Border[0 0 1 ]/H/I/C[0 1 1]}%
   user{/Subtype/Link/A<</Type/Action/S/URI/URI(#1)>>}%
  \relax
 }%
 \providecommand\@@endlink[0]{\pdfendlink}%
}%
\providecommand \url  [0]{\begingroup\@sanitize \@url }%
\providecommand \@url [1]{\endgroup\@href {#1}{\urlprefix}}%
\providecommand \urlprefix [0]{URL }%
\providecommand \Eprint[0]{\href }%
\@ifxundefined \urlstyle {%
  \providecommand \doi [1]{doi:\discretionary{}{}{}#1}%
}{%
  \providecommand \doi [0]{doi:\discretionary{}{}{}\begingroup
  \urlstyle{rm}\Url }%
}%
\providecommand \doibase [0]{http://dx.doi.org/}%
\providecommand \Doi[1]{\href{\doibase#1}}%
\providecommand \bibAnnote [3]{%
  \BibitemShut{#1}%
  \begin{quotation}\noindent
    \textsc{Key:}\ #2\\\textsc{Annotation:}\ #3%
  \end{quotation}%
}%
\providecommand \bibAnnoteFile [2]{%
  \IfFileExists{#2}{\bibAnnote {#1} {#2} {\input{#2}}}{}%
}%
\providecommand \typeout [0]{\immediate \write \m@ne }%
\providecommand \selectlanguage [0]{\@gobble}%
\providecommand \bibinfo [0]{\@secondoftwo}%
\providecommand \bibfield [0]{\@secondoftwo}%
\providecommand \translation [1]{[#1]}%
\providecommand \BibitemOpen[0]{}%
\providecommand \bibitemStop [0]{}%
\providecommand \bibitemNoStop [0]{.\EOS\space}%
\providecommand \EOS [0]{\spacefactor3000\relax}%
\providecommand \BibitemShut [1]{\csname bibitem#1\endcsname}%
%</preamble>
\bibitem{Schiff1963}%
  \BibitemOpen
  \bibfield{author}{%
  \bibinfo {author} {\bibfnamefont{L.~I.}\ \bibnamefont{Schiff}},\ }%
  \bibfield{journal}{%
  \Doi{10.1103/PhysRev.132.2194}{\bibinfo {journal} {Physical Review}}\ }%
  \textbf{\bibinfo {volume} {132}},\ \bibinfo {pages} {2194} (\bibinfo {year}
  {1963})%
  \bibAnnoteFile{NoStop}{Schiff1963}%
\bibitem{Schwinger1951}%
  \BibitemOpen
  \bibfield{author}{%
  \bibinfo {author} {\bibfnamefont{J.}~\bibnamefont{Schwinger}},\ }%
  \bibfield{journal}{%
  \Doi{10.1103/PhysRev.82.914}{\bibinfo {journal} {Physical Review}}\ }%
  \textbf{\bibinfo {volume} {82}},\ \bibinfo {pages} {914} (\bibinfo {year}
  {1951})%
  \bibAnnoteFile{NoStop}{Schwinger1951}%
\bibitem{Christenson1964}%
  \BibitemOpen
  \bibfield{author}{%
  \bibinfo {author} {\bibfnamefont{J.~H.}\ \bibnamefont{Christenson}}, \bibinfo
  {author} {\bibfnamefont{J.~W.}\ \bibnamefont{Cronin}}, \bibinfo {author}
  {\bibfnamefont{V.~L.}\ \bibnamefont{Fitch}},\ and\ \bibinfo {author}
  {\bibfnamefont{R.}~\bibnamefont{Turlay}},\ }%
  \bibfield{journal}{%
  \Doi{10.1103/PhysRevLett.13.138}{\bibinfo {journal} {Physical Review
  Letters}}\ }%
  \textbf{\bibinfo {volume} {13}},\ \bibinfo {pages} {138} (\bibinfo {year}
  {1964})%
  \bibAnnoteFile{NoStop}{Christenson1964}%
\bibitem{Balazs2005}%
  \BibitemOpen
  \bibfield{author}{%
  \bibinfo {author} {\bibfnamefont{C.}~\bibnamefont{Balazs}}, \bibinfo {author}
  {\bibfnamefont{M.}~\bibnamefont{Carena}}, \bibinfo {author}
  {\bibfnamefont{A.}~\bibnamefont{Menon}}, \bibinfo {author}
  {\bibfnamefont{D.~E.}\ \bibnamefont{Morrissey}},\ and\ \bibinfo {author}
  {\bibfnamefont{C.~E.~M.}\ \bibnamefont{Wagner}},\ }%
  \bibfield{journal}{%
  \Doi{10.1103/PhysRevD.71.075002}{\bibinfo {journal} {Physical Review D
  (Particles and Fields)}}\ }%
  \textbf{\bibinfo {volume} {71}},\ \bibinfo {pages} {075002} (\bibinfo {year}
  {2005})%
  \bibAnnoteFile{NoStop}{Balazs2005}%
\bibitem{Pospelov2005}%
  \BibitemOpen
  \bibfield{author}{%
  \bibinfo {author} {\bibfnamefont{M.}~\bibnamefont{Pospelov}}\ and\ \bibinfo
  {author} {\bibfnamefont{A.}~\bibnamefont{Ritz}},\ }%
  \bibfield{journal}{%
  \Doi{10.1016/j.aop.2005.04.002}{\bibinfo {journal} {Annals of Physics}}\ }%
  \textbf{\bibinfo {volume} {318}},\ \bibinfo {pages} {119} (\bibinfo {year}
  {2005}),\ ISSN \bibinfo {issn} {0003-4916}%
  \bibAnnoteFile{NoStop}{Pospelov2005}%
\bibitem{Booth1993}%
  \BibitemOpen
  \bibfield{author}{%
  \bibinfo {author} {\bibfnamefont{M.}~\bibnamefont{Booth}},\ }%
  \bibfield{journal}{%
  \bibinfo {journal} {ArXiv High Energy Physics - Phenomenology e-prints}}%
   (\bibinfo {year} {1993}),\
  \Eprint{http://arxiv.org/abs/arXiv:hep-ph/9301293}{arXiv:hep-ph/9301293}%
  \bibAnnoteFile{NoStop}{Booth1993}%
\bibitem{Khirplovich1991}%
  \BibitemOpen
  \bibfield{author}{%
  \bibinfo {author} {\bibfnamefont{M.}~\bibnamefont{Pospelov}}\ and\ \bibinfo
  {author} {\bibfnamefont{I.~B.}\ \bibnamefont{Khripolovich}},\ }%
  \bibfield{journal}{%
  \bibinfo {journal} {Sov. J. Nucl. Phys.}\ }%
  \textbf{\bibinfo {volume} {53}},\ \bibinfo {pages} {638} (\bibinfo {year}
  {1991})%
  \bibAnnoteFile{NoStop}{Khirplovich1991}%
\bibitem{Bernreuther1991}%
  \BibitemOpen
  \bibfield{author}{%
  \bibinfo {author} {\bibfnamefont{W.}~\bibnamefont{Bernreuther}}\ and\
  \bibinfo {author} {\bibfnamefont{M.}~\bibnamefont{Suzuki}},\ }%
  \bibfield{journal}{%
  \Doi{10.1103/RevModPhys.63.313}{\bibinfo {journal} {Reviews of Modern
  Physics}}\ }%
  \textbf{\bibinfo {volume} {63}},\ \bibinfo {pages} {313} (\bibinfo {month}
  {Apr}\ \bibinfo {year} {1991})%
  \bibAnnoteFile{NoStop}{Bernreuther1991}%
\bibitem{Hoogeveen1990}%
  \BibitemOpen
  \bibfield{author}{%
  \bibinfo {author} {\bibfnamefont{F.}~\bibnamefont{Hoogeveen}},\ }%
  \bibfield{journal}{%
  \Doi{10.1016/0550-3213(90)90182-D}{\bibinfo {journal} {Nuclear Physics B}}\
  }%
  \textbf{\bibinfo {volume} {341}},\ \bibinfo {pages} {322} (\bibinfo {year}
  {1990})%
  \bibAnnoteFile{NoStop}{Hoogeveen1990}%
\bibitem{Hudson2011}%
  \BibitemOpen
  \bibfield{author}{%
  \bibinfo {author} {\bibfnamefont{J.~J.}\ \bibnamefont{Hudson}}, \bibinfo
  {author} {\bibfnamefont{D.~M.}\ \bibnamefont{Kara}}, \bibinfo {author}
  {\bibfnamefont{I.~J.}\ \bibnamefont{Smallman}}, \bibinfo {author}
  {\bibfnamefont{B.~E.}\ \bibnamefont{Sauer}}, \bibinfo {author}
  {\bibfnamefont{M.~R.}\ \bibnamefont{Tarbutt}},\ and\ \bibinfo {author}
  {\bibfnamefont{E.~A.}\ \bibnamefont{Hinds}},\ }%
  \bibfield{journal}{%
  \Doi{10.1038/nature10104}{\bibinfo {journal} {Nature}}\ }%
  \textbf{\bibinfo {volume} {473}},\ \bibinfo {pages} {493} (\bibinfo {year}
  {2011})%
  \bibAnnoteFile{NoStop}{Hudson2011}%
\bibitem{Regan2002}%
  \BibitemOpen
  \bibfield{author}{%
  \bibinfo {author} {\bibfnamefont{B.~C.}\ \bibnamefont{Regan}}, \bibinfo
  {author} {\bibfnamefont{E.~D.}\ \bibnamefont{Commins}}, \bibinfo {author}
  {\bibfnamefont{C.~J.}\ \bibnamefont{Schmidt}},\ and\ \bibinfo {author}
  {\bibfnamefont{D.}~\bibnamefont{DeMille}},\ }%
  \bibfield{journal}{%
  \Doi{10.1103/PhysRevLett.88.071805}{\bibinfo {journal} {Phys. Rev. Lett.}}\
  }%
  \textbf{\bibinfo {volume} {88}},\ \bibinfo {pages} {071805} (\bibinfo {month}
  {Feb}\ \bibinfo {year} {2002})%
  \bibAnnoteFile{NoStop}{Regan2002}%
\bibitem{Griffith2009}%
  \BibitemOpen
  \bibfield{author}{%
  \bibinfo {author} {\bibfnamefont{W.~C.}\ \bibnamefont{Griffith}}, \bibinfo
  {author} {\bibfnamefont{M.~D.}\ \bibnamefont{Swallows}}, \bibinfo {author}
  {\bibfnamefont{T.~H.}\ \bibnamefont{Loftus}}, \bibinfo {author}
  {\bibfnamefont{M.~V.}\ \bibnamefont{Romalis}}, \bibinfo {author}
  {\bibfnamefont{B.~R.}\ \bibnamefont{Heckel}},\ and\ \bibinfo {author}
  {\bibfnamefont{E.~N.}\ \bibnamefont{Fortson}},\ }%
  \bibfield{journal}{%
  \Doi{10.1103/PhysRevLett.102.101601}{\bibinfo {journal} {Phys. Rev. Lett.}}\
  }%
  \textbf{\bibinfo {volume} {102}},\ \bibinfo {pages} {101601} (\bibinfo {year}
  {2009})%
  \bibAnnoteFile{NoStop}{Griffith2009}%
\bibitem{KhriplovichLamoreaux}%
  \BibitemOpen
  \bibfield{author}{%
  \bibinfo {author} {\bibfnamefont{I.~B.}\ \bibnamefont{Khriplovich}}\ and\
  \bibinfo {author} {\bibfnamefont{S.~K.}\ \bibnamefont{Lamoreaux}},\ }%
  \emph{\bibinfo {title} {CP Violation without Strangeness}}\ (\bibinfo
  {publisher} {Springer-Verlag},\ \bibinfo {address} {Berlin},\ \bibinfo {year}
  {1997})%
  \bibAnnoteFile{NoStop}{KhriplovichLamoreaux}%
\bibitem{Commins2010}%
  \BibitemOpen
  \bibfield{author}{%
  \bibinfo {author} {\bibfnamefont{E.}~\bibnamefont{Commins}}\ and\ \bibinfo
  {author} {\bibfnamefont{D.}~\bibnamefont{DeMille}},\ }%
  \bibfield{journal}{%
  \bibinfo {journal} {Lepton Dipole Moments}}%
   (\bibinfo {year} {2010})%
  \bibAnnoteFile{NoStop}{Commins2010}%
\bibitem{Meyer2006}%
  \BibitemOpen
  \bibfield{author}{%
  \bibinfo {author} {\bibfnamefont{E.~R.}\ \bibnamefont{Meyer}}, \bibinfo
  {author} {\bibfnamefont{J.~L.}\ \bibnamefont{Bohn}},\ and\ \bibinfo {author}
  {\bibfnamefont{M.~P.}\ \bibnamefont{Deskevich}},\ }%
  \bibfield{journal}{%
  \Doi{10.1103/PhysRevA.73.062108}{\bibinfo {journal} {Phys. Rev. A}}\ }%
  \textbf{\bibinfo {volume} {73}},\ \bibinfo {pages} {062108} (\bibinfo {year}
  {2006})%
  \bibAnnoteFile{NoStop}{Meyer2006}%
\bibitem{Lee2009}%
  \BibitemOpen
  \bibfield{author}{%
  \bibinfo {author} {\bibfnamefont{J.}~\bibnamefont{Lee}}, \bibinfo {author}
  {\bibfnamefont{E.~R.}\ \bibnamefont{Meyer}}, \bibinfo {author}
  {\bibfnamefont{R.}~\bibnamefont{Paudel}}, \bibinfo {author}
  {\bibfnamefont{J.~L.}\ \bibnamefont{Bohn}},\ and\ \bibinfo {author}
  {\bibfnamefont{A.~E.}\ \bibnamefont{Leanhardt}},\ }%
  \bibfield{journal}{%
  \Doi{10.1080/09500340903349930}{\bibinfo {journal} {Journal of Modern
  Optics}}\ }%
  \textbf{\bibinfo {volume} {56}},\ \bibinfo {pages} {2005} (\bibinfo {year}
  {2009})%
  \bibAnnoteFile{NoStop}{Lee2009}%
\bibitem{Vutha2010}%
  \BibitemOpen
  \bibfield{author}{%
  \bibinfo {author} {\bibfnamefont{A.~C.}\ \bibnamefont{Vutha}}, \bibinfo
  {author} {\bibfnamefont{W.~C.}\ \bibnamefont{Campbell}}, \bibinfo {author}
  {\bibfnamefont{Y.~V.}\ \bibnamefont{Gurevich}}, \bibinfo {author}
  {\bibfnamefont{N.~R.}\ \bibnamefont{Hutzler}}, \bibinfo {author}
  {\bibfnamefont{M.}~\bibnamefont{Parsons}}, \bibinfo {author}
  {\bibfnamefont{D.}~\bibnamefont{Patterson}}, \bibinfo {author}
  {\bibfnamefont{E.}~\bibnamefont{Petrik}}, \bibinfo {author}
  {\bibfnamefont{B.}~\bibnamefont{Spaun}}, \bibinfo {author}
  {\bibfnamefont{J.~M.}\ \bibnamefont{Doyle}}, \bibinfo {author}
  {\bibfnamefont{G.}~\bibnamefont{Gabrielse}},\ and\ \bibinfo {author}
  {\bibfnamefont{D.}~\bibnamefont{DeMille}},\ }%
  \bibfield{journal}{%
  \bibinfo {journal} {Journal of Physics B: Atomic, Molecular and Optical
  Physics}\ }%
  \textbf{\bibinfo {volume} {43}},\ \bibinfo {pages} {074007} (\bibinfo {year}
  {2010})%
  \bibAnnoteFile{NoStop}{Vutha2010}%
\bibitem{Loh2011}%
  \BibitemOpen
  \bibfield{author}{%
  \bibinfo {author} {\bibfnamefont{H.}~\bibnamefont{Loh}}, \bibinfo {author}
  {\bibfnamefont{J.}~\bibnamefont{Wang}}, \bibinfo {author}
  {\bibfnamefont{M.}~\bibnamefont{Grau}}, \bibinfo {author}
  {\bibfnamefont{T.~S.}\ \bibnamefont{Yahn}}, \bibinfo {author}
  {\bibfnamefont{R.~W.}\ \bibnamefont{Field}}, \bibinfo {author}
  {\bibfnamefont{C.~H.}\ \bibnamefont{Greene}},\ and\ \bibinfo {author}
  {\bibfnamefont{E.~A.}\ \bibnamefont{Cornell}},\ }%
  \bibfield{journal}{%
  \Doi{10.1063/1.3652333}{\bibinfo {journal} {The Journal of Chemical
  Physics}}\ }%
  \textbf{\bibinfo {volume} {135}},\ \bibinfo {pages} {154308} (\bibinfo {year}
  {2011})%
  \bibAnnoteFile{NoStop}{Loh2011}%
\bibitem{Grau2012}%
  \BibitemOpen
  \bibfield{author}{%
  \bibinfo {author} {\bibfnamefont{M.}~\bibnamefont{Grau}}, \bibinfo {author}
  {\bibfnamefont{A.~E.}\ \bibnamefont{Leanhardt}}, \bibinfo {author}
  {\bibfnamefont{H.}~\bibnamefont{Loh}}, \bibinfo {author}
  {\bibfnamefont{L.~C.}\ \bibnamefont{Sinclair}}, \bibinfo {author}
  {\bibfnamefont{R.~P.}\ \bibnamefont{Stutz}}, \bibinfo {author}
  {\bibfnamefont{T.~S.}\ \bibnamefont{Yahn}},\ and\ \bibinfo {author}
  {\bibfnamefont{E.~A.}\ \bibnamefont{Cornell}},\ }%
  \bibfield{journal}{%
  \Doi{10.1016/j.jms.2011.12.006}{\bibinfo {journal} {Journal of Molecular
  Spectroscopy}}\ }%
  \textbf{\bibinfo {volume} {272}},\ \bibinfo {pages} {32} (\bibinfo {year}
  {2012})%
  \bibAnnoteFile{NoStop}{Grau2012}%
\bibitem{DeMille2001}%
  \BibitemOpen
  \bibfield{author}{%
  \bibinfo {author} {\bibfnamefont{D.}~\bibnamefont{DeMille}}, \bibinfo
  {author} {\bibfnamefont{F.}~\bibnamefont{Bay}}, \bibinfo {author}
  {\bibfnamefont{S.}~\bibnamefont{Bickman}}, \bibinfo {author}
  {\bibfnamefont{D.}~\bibnamefont{Kawall}}, \bibinfo {author}
  {\bibfnamefont{L.}~\bibnamefont{Hunter}}, \bibinfo {author}
  {\bibfnamefont{D.}~\bibnamefont{Krause~Jr.}}, \bibinfo {author}
  {\bibfnamefont{S.}~\bibnamefont{Maxwell}},\ and\ \bibinfo {author}
  {\bibfnamefont{K.}~\bibnamefont{Ulmer}},\ }%
  \bibfield{booktitle}{%
  \emph{\bibinfo {booktitle} {Art and Symmetry in Experimental Physics}},\ }%
  \bibinfo {series} {AIP Conf. Proc.}\ \textbf{\bibinfo {volume} {596}},\
  \bibinfo {pages} {72} (\bibinfo {year} {2001})%
  \bibAnnoteFile{NoStop}{DeMille2001}%
\bibitem{Eckel2012}%
  \BibitemOpen
  \bibfield{author}{%
  \bibinfo {author} {\bibfnamefont{S.}~\bibnamefont{Eckel}}, \bibinfo {author}
  {\bibfnamefont{A.~O.}\ \bibnamefont{Sushkov}},\ and\ \bibinfo {author}
  {\bibfnamefont{S.~K.}\ \bibnamefont{Lamoreaux}},\ }%
  \bibfield{journal}{%
  \Doi{10.1103/PhysRevLett.109.193003}{\bibinfo {journal} {Phys. Rev. Lett.}}\
  }%
  \textbf{\bibinfo {volume} {109}},\ \bibinfo {pages} {193003} (\bibinfo
  {month} {Nov}\ \bibinfo {year} {2012})%
  \bibAnnoteFile{NoStop}{Eckel2012}%
\bibitem{Rushchanskii2010}%
  \BibitemOpen
  \bibfield{author}{%
  \bibinfo {author} {\bibfnamefont{K.~Z.}\ \bibnamefont{Rushchanskii}},
  \bibinfo {author} {\bibfnamefont{S.}~\bibnamefont{Kamba}}, \bibinfo {author}
  {\bibfnamefont{V.}~\bibnamefont{Goian}}, \bibinfo {author}
  {\bibfnamefont{P.}~\bibnamefont{Van{\v{e}}k}}, \bibinfo {author}
  {\bibfnamefont{M.}~\bibnamefont{Savinov}}, \bibinfo {author}
  {\bibfnamefont{J.}~\bibnamefont{Prokle{\v{s}}ka}}, \bibinfo {author}
  {\bibfnamefont{D.}~\bibnamefont{Nuzhnyy}}, \bibinfo {author}
  {\bibfnamefont{K.}~\bibnamefont{Kn{\'{i}}{\v{z}}ek}}, \bibinfo {author}
  {\bibfnamefont{F.}~\bibnamefont{Laufek}}, \bibinfo {author}
  {\bibfnamefont{S.}~\bibnamefont{Eckel}}, \bibinfo {author}
  {\bibfnamefont{S.~K.}\ \bibnamefont{Lamoreaux}}, \bibinfo {author}
  {\bibfnamefont{A.~O.}\ \bibnamefont{Sushkov}}, \bibinfo {author}
  {\bibfnamefont{M.}~\bibnamefont{Le{\v{z}}ai{\'c}}},\ and\ \bibinfo {author}
  {\bibfnamefont{N.~A.}\ \bibnamefont{Spaldin}},\ }%
  \bibfield{journal}{%
  \Doi{10.1038/nmat2799}{
  \bibinfo {journal} {Nature Materials}\ }}%
  \textbf{\bibinfo {volume} {9}},\ \bibinfo {pages} {649} (\bibinfo {month}
  {08}\ \bibinfo {year} {2010}) %
  \bibAnnoteFile{NoStop}{Rushchanskii2010}%
\bibitem{Weiss2003}%
  \BibitemOpen
  \bibfield{author}{%
  \bibinfo {author} {\bibfnamefont{F.~F.}\ \bibnamefont{D.~Weiss}}\ and\
  \bibinfo {author} {\bibfnamefont{J.}~\bibnamefont{Chen}},\ }%
  \bibfield{journal}{%
  \bibinfo {journal} {Bull. Am. Phys. Soc.}\ }%
  \textbf{\bibinfo {volume} {APR03}},\ \bibinfo {pages} {J1.008} (\bibinfo
  {year} {2003})%
  \bibAnnoteFile{NoStop}{Weiss2003}%
\bibitem{Bickman2009}%
  \BibitemOpen
  \bibfield{author}{%
  \bibinfo {author} {\bibfnamefont{S.}~\bibnamefont{Bickman}}, \bibinfo
  {author} {\bibfnamefont{P.}~\bibnamefont{Hamilton}}, \bibinfo {author}
  {\bibfnamefont{Y.}~\bibnamefont{Jiang}},\ and\ \bibinfo {author}
  {\bibfnamefont{D.}~\bibnamefont{DeMille}},\ }%
  \bibfield{journal}{%
  \Doi{10.1103/PhysRevA.80.023418}{\bibinfo {journal} {Phys. Rev. A}}\ }%
  \textbf{\bibinfo {volume} {80}},\ \bibinfo {pages} {023418} (\bibinfo {year}
  {2009})%
  \bibAnnoteFile{NoStop}{Bickman2009}%
\bibitem{Hamilton2010}%
  \BibitemOpen
  \bibfield{author}{%
  \bibinfo {author} {\bibfnamefont{P.}~\bibnamefont{Hamilton}},\ }%
  \emph{\bibinfo {title} {Preliminary results in the search for the electron
  electric dipole moment in PbO$^*$}},\ Ph.D. thesis,\ \bibinfo {school} {Yale
  University}, \bibinfo {address} {P.O. Box 208120, New Haven, CT 06520-8120}
  (\bibinfo {year} {2010})%
  \bibAnnoteFile{NoStop}{Hamilton2010}%
\bibitem{DeMille2000}%
  \BibitemOpen
  \bibfield{author}{%
  \bibinfo {author} {\bibfnamefont{D.}~\bibnamefont{DeMille}}, \bibinfo
  {author} {\bibfnamefont{F.}~\bibnamefont{Bay}}, \bibinfo {author}
  {\bibfnamefont{S.}~\bibnamefont{Bickman}}, \bibinfo {author}
  {\bibfnamefont{D.}~\bibnamefont{Kawall}}, \bibinfo {author}
  {\bibfnamefont{D.}~\bibnamefont{Krause}}, \bibinfo {author}
  {\bibfnamefont{S.~E.}\ \bibnamefont{Maxwell}},\ and\ \bibinfo {author}
  {\bibfnamefont{L.~R.}\ \bibnamefont{Hunter}},\ }%
  \bibfield{journal}{%
  \Doi{10.1103/PhysRevA.61.052507}{\bibinfo {journal} {Phys. Rev. A}}\ }%
  \textbf{\bibinfo {volume} {61}},\ \bibinfo {pages} {052507} (\bibinfo {year}
  {2000})%
  \bibAnnoteFile{NoStop}{DeMille2000}%
\bibitem{Hunter2002}%
  \BibitemOpen
  \bibfield{author}{%
  \bibinfo {author} {\bibfnamefont{L.~R.}\ \bibnamefont{Hunter}}, \bibinfo
  {author} {\bibfnamefont{S.~E.}\ \bibnamefont{Maxwell}}, \bibinfo {author}
  {\bibfnamefont{K.~A.}\ \bibnamefont{Ulmer}}, \bibinfo {author}
  {\bibfnamefont{N.~D.}\ \bibnamefont{Charney}}, \bibinfo {author}
  {\bibfnamefont{S.~K.}\ \bibnamefont{Peck}}, \bibinfo {author}
  {\bibfnamefont{D.}~\bibnamefont{Krause}}, \bibinfo {author}
  {\bibfnamefont{S.}~\bibnamefont{Ter-Avetisyan}},\ and\ \bibinfo {author}
  {\bibfnamefont{D.}~\bibnamefont{DeMille}},\ }%
  \bibfield{journal}{%
  \Doi{10.1103/PhysRevA.65.030501}{\bibinfo {journal} {Phys. Rev. A}}\ }%
  \textbf{\bibinfo {volume} {65}},\ \bibinfo {pages} {030501} (\bibinfo {year}
  {2002})%
  \bibAnnoteFile{NoStop}{Hunter2002}%
\bibitem{Kozlov2002}%
  \BibitemOpen
  \bibfield{author}{%
  \bibinfo {author} {\bibfnamefont{M.~G.}\ \bibnamefont{Kozlov}}\ and\ \bibinfo
  {author} {\bibfnamefont{D.}~\bibnamefont{DeMille}},\ }%
  \bibfield{journal}{%
  \Doi{10.1103/PhysRevLett.89.133001}{\bibinfo {journal} {Phys. Rev. Lett.}}\
  }%
  \textbf{\bibinfo {volume} {89}},\ \bibinfo {pages} {133001} (\bibinfo {year}
  {2002})%
  \bibAnnoteFile{NoStop}{Kozlov2002}%
\bibitem{Kawall2004}%
  \BibitemOpen
  \bibfield{author}{%
  \bibinfo {author} {\bibfnamefont{D.}~\bibnamefont{Kawall}}, \bibinfo {author}
  {\bibfnamefont{F.}~\bibnamefont{Bay}}, \bibinfo {author}
  {\bibfnamefont{S.}~\bibnamefont{Bickman}}, \bibinfo {author}
  {\bibfnamefont{Y.}~\bibnamefont{Jiang}},\ and\ \bibinfo {author}
  {\bibfnamefont{D.}~\bibnamefont{DeMille}},\ }%
  \bibfield{journal}{%
  \Doi{10.1103/PhysRevLett.92.133007}{\bibinfo {journal} {Phys. Rev. Lett.}}\
  }%
  \textbf{\bibinfo {volume} {92}},\ \bibinfo {pages} {133007} (\bibinfo {year}
  {2004})%
  \bibAnnoteFile{NoStop}{Kawall2004}%
\bibitem{Petrov2005}%
  \BibitemOpen
  \bibfield{author}{%
  \bibinfo {author} {\bibfnamefont{A.~N.}\ \bibnamefont{Petrov}}, \bibinfo
  {author} {\bibfnamefont{A.~V.}\ \bibnamefont{Titov}}, \bibinfo {author}
  {\bibfnamefont{T.~A.}\ \bibnamefont{Isaev}}, \bibinfo {author}
  {\bibfnamefont{N.~S.}\ \bibnamefont{Mosyagin}},\ and\ \bibinfo {author}
  {\bibfnamefont{D.}~\bibnamefont{DeMille}},\ }%
  \bibfield{journal}{%
  \Doi{10.1103/PhysRevA.72.022505}{\bibinfo {journal} {Phys. Rev. A}}\ }%
  \textbf{\bibinfo {volume} {72}},\ \bibinfo {pages} {022505} (\bibinfo {year}
  {2005})%
  \bibAnnoteFile{NoStop}{Petrov2005}%
\bibitem{Martin1988}%
  \BibitemOpen
  \bibfield{author}{%
  \bibinfo {author} {\bibfnamefont{F.}~\bibnamefont{Martin}}, \bibinfo {author}
  {\bibfnamefont{R.}~\bibnamefont{Bacis}}, \bibinfo {author}
  {\bibfnamefont{J.}~\bibnamefont{Verg{\`{e}}s}}, \bibinfo {author}
  {\bibfnamefont{J.}~\bibnamefont{Bachar}},\ and\ \bibinfo {author}
  {\bibfnamefont{S.}~\bibnamefont{Rosenwaks}},\ }%
  \bibfield{journal}{%
  \Doi{10.1016/0584-8539(88)80005-9}{\bibinfo {journal} {Spectrochimica Acta
  Part A: Molecular Spectroscopy}}\ }%
  \textbf{\bibinfo {volume} {44}},\ \bibinfo {pages} {889} (\bibinfo {year}
  {1988})%
  \bibAnnoteFile{NoStop}{Martin1988}%
\bibitem{Bailey1978}%
  \BibitemOpen
  \bibfield{author}{%
  \bibinfo {author} {\bibfnamefont{F.~P.}\ \bibnamefont{Bailey}}\ and\ \bibinfo
  {author} {\bibfnamefont{K.~J.~T.}\ \bibnamefont{Black}},\ }%
  \bibfield{journal}{%
  \Doi{10.1007/BF00544700}{\bibinfo {journal} {Journal of Materials Science}}\
  }%
  \textbf{\bibinfo {volume} {13}},\ \bibinfo {pages} {1045} (\bibinfo {year}
  {1978})%
  \bibAnnoteFile{NoStop}{Bailey1978}%
\bibitem{Lopatina2006}%
  \BibitemOpen
  \bibfield{author}{%
  \bibinfo {author} {\bibfnamefont{S.~I.}\ \bibnamefont{Lopatina}}, \bibinfo
  {author} {\bibfnamefont{I.~Y.}\ \bibnamefont{Mittovab}}, \bibinfo {author}
  {\bibfnamefont{F.~S.}\ \bibnamefont{Gerasimovb}}, \bibinfo {author}
  {\bibfnamefont{S.~M.}\ \bibnamefont{Shugurova}}, \bibinfo {author}
  {\bibfnamefont{V.~F.}\ \bibnamefont{Kostryukovb}},\ and\ \bibinfo {author}
  {\bibfnamefont{S.~M.}\ \bibnamefont{Skorokhodovab}},\ }%
  \bibfield{journal}{%
  \Doi{10.1134/S0036023606100214}{\bibinfo {journal} {Russ. J. Inorg. Chem.}}\
  }%
  \textbf{\bibinfo {volume} {51}},\ \bibinfo {pages} {1646} (\bibinfo {year}
  {2006})%
  \bibAnnoteFile{NoStop}{Lopatina2006}%
\bibitem{Popovic1997}%
  \BibitemOpen
  \bibfield{author}{%
  \bibinfo {author} {\bibfnamefont{A.}~\bibnamefont{Popovi{\v{c}}}}, \bibinfo
  {author} {\bibfnamefont{A.}~\bibnamefont{Lesar}}, \bibinfo {author}
  {\bibfnamefont{M.}~\bibnamefont{Gu{\v{c}}ek}},\ and\ \bibinfo {author}
  {\bibfnamefont{L.}~\bibnamefont{Bencze}},\ }%
  \bibfield{journal}{%
  \Doi{10.1002/(SICI)1097-0231(199703)11:5<459::AID-RCM889>3.0.CO;2-G}{\bibinfo
  {journal} {Rapid Communications in Mass Spectrometry}}\ }%
  \textbf{\bibinfo {volume} {11}},\ \bibinfo {pages} {459} (\bibinfo {year}
  {1997})%
  \bibAnnoteFile{NoStop}{Popovic1997}%
\bibitem{Schwettmann2007}%
  \BibitemOpen
  \bibfield{author}{%
  \bibinfo {author} {\bibfnamefont{A.}~\bibnamefont{Schwettmann}}, \bibinfo
  {author} {\bibfnamefont{C.}~\bibnamefont{McGuffey}}, \bibinfo {author}
  {\bibfnamefont{S.}~\bibnamefont{Chauhan}}, \bibinfo {author}
  {\bibfnamefont{K.~R.}\ \bibnamefont{Overstreet}},\ and\ \bibinfo {author}
  {\bibfnamefont{J.~P.}\ \bibnamefont{Shaffer}},\ }%
  \bibfield{journal}{%
  \Doi{10.1364/AO.46.001310}{\bibinfo {journal} {Appl. Opt.}}\ }%
  \textbf{\bibinfo {volume} {46}},\ \bibinfo {pages} {1310} (\bibinfo {month}
  {Mar}\ \bibinfo {year} {2007})%
  \bibAnnoteFile{NoStop}{Schwettmann2007}%
\bibitem{Drell1979}%
  \BibitemOpen
  \bibfield{author}{%
  \bibinfo {author} {\bibfnamefont{P.}~\bibnamefont{Drell}}\ and\ \bibinfo
  {author} {\bibfnamefont{S.}~\bibnamefont{Chu}},\ }%
  \bibfield{journal}{%
  \Doi{10.1016/0030-4018(79)90335-3}{\bibinfo {journal} {Optics
  Communications}}\ }%
  \textbf{\bibinfo {volume} {28}},\ \bibinfo {pages} {343} (\bibinfo {year}
  {1979})%
  \bibAnnoteFile{NoStop}{Drell1979}%
\bibitem{Oppenheim2010}%
  \BibitemOpen
  \bibfield{author}{%
  \bibinfo {author} {\bibfnamefont{A.~V.}\ \bibnamefont{Oppenheim}}\ and\
  \bibinfo {author} {\bibfnamefont{R.~W.}\ \bibnamefont{Schafer}},\ }%
  \emph{\bibinfo {title} {Discrete-time signal processing}},\ Prentice-Hall
  signal processing series\ (\bibinfo {publisher} {Prentice Hall},\ \bibinfo
  {year} {2010})\ ISBN \bibinfo {isbn} {9780131988422}%
  \bibAnnoteFile{NoStop}{Oppenheim2010}%
\bibitem{Jiang2008}%
  \BibitemOpen
  \bibfield{author}{%
  \bibinfo {author} {\bibfnamefont{Y.}~\bibnamefont{Jiang}},\ }%
  \emph{\bibinfo {title} {Progress toward searching for electron electric
  dipole moment using lead oxide}},\ Ph.D. thesis,\ \bibinfo {school} {Yale
  University}, \bibinfo {address} {P.O. Box 208120, New Haven, CT 06520-8120}
  (\bibinfo {year} {2008})%
  \bibAnnoteFile{NoStop}{Jiang2008}%
\bibitem{Maxwell1904}%
  \BibitemOpen
  \bibfield{author}{%
  \bibinfo {author} {\bibfnamefont{J.~C.}\ \bibnamefont{Maxwell}}\ and\
  \bibinfo {author} {\bibfnamefont{J.~J.}\ \bibnamefont{Thompson}},\ }%
  \emph{\bibinfo {title} {A Treatise on Electricity and Magnetism}},\ \bibinfo
  {series} {Clarendon Press series}\ No.\ \bibinfo {number} {v. 1}\ (\bibinfo
  {publisher} {Clarendon},\ \bibinfo {year} {1904})%
  \bibAnnoteFile{NoStop}{Maxwell1904}%
\bibitem{Bevington2003}%
  \BibitemOpen
  \bibfield{author}{%
  \bibinfo {author} {\bibfnamefont{P.~R.}\ \bibnamefont{Bevington}}\ and\
  \bibinfo {author} {\bibfnamefont{D.~K.}\ \bibnamefont{Robinson}},\ }%
  \emph{\bibinfo {title} {Data reduction and error analysis for the physical
  sciences}},\ McGraw-Hill Higher Education\ (\bibinfo {publisher}
  {McGraw-Hill},\ \bibinfo {year} {2003})\ ISBN \bibinfo {isbn}
  {9780072472271}%
  \bibAnnoteFile{NoStop}{Bevington2003}%
\bibitem{KhriplovichLamoreaux1997}%
  \BibitemOpen
  \bibfield{author}{%
  \bibinfo {author} {\bibfnamefont{I.~B.}\ \bibnamefont{Khriplovich}}\ and\
  \bibinfo {author} {\bibfnamefont{S.~K.}\ \bibnamefont{Lamoreaux}},\ }%
  \emph{\bibinfo {title} {CP Violation without Strangeness}}\ (\bibinfo
  {publisher} {Springer-Verlag},\ \bibinfo {address} {Berlin},\ \bibinfo {year}
  {1997})%
  \bibAnnoteFile{NoStop}{KhriplovichLamoreaux1997}%
\bibitem{Vutha2011}%
  \BibitemOpen
  \bibfield{author}{%
  \bibinfo {author} {\bibfnamefont{A.~C.}\ \bibnamefont{Vutha}}, \bibinfo
  {author} {\bibfnamefont{B.}~\bibnamefont{Spaun}}, \bibinfo {author}
  {\bibfnamefont{Y.~V.}\ \bibnamefont{Gurevich}}, \bibinfo {author}
  {\bibfnamefont{N.~R.}\ \bibnamefont{Hutzler}}, \bibinfo {author}
  {\bibfnamefont{E.}~\bibnamefont{Kirilov}}, \bibinfo {author}
  {\bibfnamefont{J.~M.}\ \bibnamefont{Doyle}}, \bibinfo {author}
  {\bibfnamefont{G.}~\bibnamefont{Gabrielse}},\ and\ \bibinfo {author}
  {\bibfnamefont{D.}~\bibnamefont{DeMille}},\ }%
  \bibfield{journal}{%
  \Doi{10.1103/PhysRevA.84.034502}{\bibinfo {journal} {Phys. Rev. A}}\ }%
  \textbf{\bibinfo {volume} {84}},\ \bibinfo {pages} {034502} (\bibinfo {year}
  {2011})%
  \bibAnnoteFile{NoStop}{Vutha2011}%
\end{thebibliography}
\end{document}